\documentclass[journal=jacsat,manuscript=article]{achemso}

\usepackage{chemformula} 
\usepackage[T1]{fontenc} 
\usepackage{soul}
\usepackage{xcolor}
\definecolor{mygreen}{RGB}{0, 128, 0}

\sethlcolor{lime}
\usepackage{colortbl}
\usepackage{xcolor}
\usepackage{siunitx}
\usepackage{hyperref}

\newcommand{\vb}[0]{$\mathrm{V_B^-}$ }
\newcommand{\eg}[0]{$E_\mathrm{2g}$ }

\author{Atanu Patra}
\affiliation{Julius-Maximilians-Universität Würzburg, Lehrstuhl für Technische Physik, Am Hubland, Würzburg 97074, Deutschland}
\altaffiliation{These authors contributed equally to this work}
\email{atanu.patra@uni-wuerzburg.de}
\author{Paul Konrad}
\affiliation{Julius-Maximilians-Universität Würzburg, Experimental Physics 6, Am Hubland, Würzburg 97074, Deutschland}
\altaffiliation{These authors contributed equally to this work}
\author{Andreas Sperlich}
\affiliation{Julius-Maximilians-Universität Würzburg, Experimental Physics 6, Am Hubland, Würzburg 97074, Deutschland}
\altaffiliation{Physikalisches Institut and Würzburg-Dresden Cluster of Excellence ct.qmat, Deutschland}

\author{Timur Biktagirov}
\affiliation{Universität Paderborn, Department Physik, Warburger Str. 100, 33098 Paderborn, Deutschland}
\author{Wolf Gero Schmidt}
\affiliation{Universität Paderborn, Department Physik, Warburger Str. 100, 33098 Paderborn, Deutschland}

\author{Lesley Spencer}
\affiliation{School of Mathematical and Physical Sciences, University of Technology Sydney, Ultimo, New South Wales
2007, Australia}
\author{Igor Aharonovich}
\affiliation{School of Mathematical and Physical Sciences, University of Technology Sydney, Ultimo, New South Wales
2007, Australia}
\altaffiliation{ARC Centre of Excellence for Transformative Meta-Optical Systems, University of Technology Sydney,
Ultimo, New South Wales 2007, Australia}
\author{Sven Höfling}
\affiliation{Julius-Maximilians-Universität Würzburg, Lehrstuhl für Technische Physik, Am Hubland, Würzburg 97074, Deutschland}
\altaffiliation{Physikalisches Institut and Würzburg-Dresden Cluster of Excellence ct.qmat, Deutschland}

\author{Vladimir Dyakonov}
\affiliation{Julius-Maximilians-Universität Würzburg, Experimental Physics 6, Am Hubland, Würzburg 97074, Deutschland}

\altaffiliation{Physikalisches Institut and Würzburg-Dresden Cluster of Excellence ct.qmat, Deutschland}

\email{vladimir.dyakonov@uni-wuerzburg.de}

\title[An \textsf{achemso} demo]
  {Quantifying Spin Defect Density in hBN via Raman and Photoluminescence Analysis}

\keywords{hBN,  $V_B^-$ spin defects, Raman, photoluminescence, polarization dependence}

\begin{document}

\newpage
\begin{abstract}
Negatively charged boron vacancies ($\mathrm{V_B^-}$) in hexagonal boron nitride (hBN) are emerging as promising solid-state spin qubits due to their optical accessibility, structural simplicity, and compatibility with photonic platforms. However, quantifying the density of such defects in thin hBN flakes has remained elusive, limiting progress in device integration and reproducibility. Here, we present an all-optical method to quantify $\mathrm{V_B^-}$ defect density in hBN by correlating Raman and photoluminescence (PL) signatures with irradiation fluence. We identify two defect-induced Raman modes, D1 and D2, and assign them to vibrational modes of $\mathrm{V_B^-}$ using polarization-resolved Raman measurements and density functional theory (DFT) calculations. By adapting a numerical model originally developed for graphene, we establish an empirical relationship linking Raman (D1, \eg) and PL intensities to absolute defect densities. This method is universally applicable across various irradiation types and uniquely suited for thin flakes, where conventional techniques fail. Our approach enables accurate, direct, and non-destructive quantification of spin defect densities down to \SI{e15} {defects\per \centi\meter\cubed}, offering a powerful tool for optimizing and benchmarking hBN for quantum optical applications.


\end{abstract}

\textbf{keywords:} hBN,  $\mathrm{V_B^-}$, spin defects, Raman, photoluminescence, DFT
\section{Introduction}

\textbf{H}exagonal  boron nitride (hBN) has become an indispensable part of device fabrication using van der Waals (vdW) materials, serving both as an encapsulation and an insulator due to its large band gap of $\SI{6}{\electronvolt}$ \cite{cassabois2016hexagonal}.
Furthermore, identification of optically active negatively charged boron vacancy spin defects (\vb) in hBN by Gottscholl \textit{et. al.}\cite{gottscholl2020initialization} has garnered significant interest in the field of quantum sensing as \vb offers a robust platform for quantum technologies that can operate at room temperature. 
Specifically, it has been shown that \vb centers are highly sensitive to their environment \cite{gottscholl2020initialization,gottscholl2021room,chejanovsky2021single,haykal2022decoherence}, making them ideal quantum sensors for magnetic fields \cite{huang2022wide,rizzato2023extending}, temperature, and pressure \cite{gottscholl2021spin}.
In general, due to their reduced dimensionality, spin defects in two-dimensional (2D) van der Waals (vdW) materials offer unique advantages over their counterparts in three-dimensional (3D) systems, such as nitrogen-vacancy centers in diamond and spin defects in silicon carbide. In the case of hBN, this 2D nature allows for the creation of stable spin defects near the sample surface, significantly enhancing the precision of quantum sensing.\cite{gottscholl2020initialization,chejanovsky2021single} 
Moreover, the atomic thinness, stability, and chemical inertness of hBN provide a scalable platform for integrating spin defects into various hybrid photonic devices \cite{froch2021coupling,qian2022unveiling,moon2024fiber} and even for use in studying complex charge dynamics in aqueous environments \cite{comtet2020direct,comtet2021anomalous}. 
Such applications of hBN largely depend on the precise control and distribution of spin defects in it. 
Although defect generation can be achieved by various types of irradiaiton like focused ion beams \cite{kianinia2020generation}, neutrons \cite{gottscholl2020initialization, li2021defect}, laser writing \cite{gao2021femtosecond} and electrons \cite{murzakhanov2021creation}, the absolute spin-defect density is greatly dependent on irradiation parameters and sample properties. 

The emitted photoluminescence (PL) then depends on the type of defects present in the system \cite{liu2022spin}. 
The PL of \vb has a broad, featureless spectrum in the near infrared around 850 nm. So far, the zero phonon line has not been resolved yet but was estimated to be at 773 nm.\cite{qian2022unveiling}
The electronic dipole moment linked to a defect influences its PL. Furthermore, we also investigate the defect-related vibrational modes and hence correlation to defect PL.
In this regard, Raman spectroscopy is a valuable tool in the understanding of the vibrational modes in 2D materials. In particular, Raman spectroscopy of graphene has been instrumental in the study of its crystal structure and defects. 
Both, graphene and hBN have a hexagonal crystal structure with $D_\mathrm{6h}$ and $D_\mathrm{3h}$ point-group symmetry respectively. The main difference is the unit cell composition. Graphene consists of carbon atoms, whereas hBN contains boron and nitrogen, thus reducing symmetry.
The characteristic Raman peak in graphene is at 1585 cm$^{-1}$, which is a plane high frequency \eg and, additionally, a  defect activated mode, D peak is at 1350 cm$^{-1}$\cite{tuinstra1970raman,ferrari2013raman}. 
Several works have shown the quantification of the crystal size ($L_\mathrm{a}$) in graphene, and thereby the defect density from the relative contribution of the \eg and D peak in nanographites \cite{tuinstra1970raman, ferrari2000interpretation, canccado2006general, zickler2006reconsideration, canccado2007measuring}, amorphous carbons, carbon nanotubes and graphene \cite{lucchese2010quantifying, canccado2011quantifying}.
Similar to graphene, recent studies have identified \cite{li2021defect,venturi2024selective,ren2023creation,carbone2025quantifying} a defect-related Raman peak in hBN. Vibrational modes of several prominent defects in hBN have also been addressed theoretically  \cite{linderalv2021vibrational}.
Although methods such as electron paramagnetic resonance (EPR) \cite{gottscholl2021spin} and single defect counting by PL mapping exist for an estimation of emitter density, they are hardly applicable to the low absolute number of defects found in thin flakes that also have a low photon count rate. 
Furthermore, a standardized protocol for direct determination of spin-defect density solely from spectral data could not be established yet.


In this work, we have investigated two emerging Raman modes, labeled as D1 and D2 following the convention from the graphene literature, alongside the well-established \eg mode in hBN. 
We find that these newly identified Raman modes are associated with \vb defects, which emit a distinctive PL signal in the near infrared under laser excitation, and support this assignment using density functional theory (DFT) calculations.
Both, D1 and PL intensities, increase with laser excitation energy ($E_\mathrm{L}$), providing insights into the optical absorption and intrinsic properties of \vb spin defects.
Polarization-dependent Raman measurements were employed to understand the symmetry and origin of these modes. 
We generated \vb defects by focused nitrogen ion beam irradiation in hBN with varying fluence and find that the intensities of Raman modes and PL signal correlate with the fluence, and hence with defect density. 
By analyzing the dependence of area of the D1 and \eg Raman modes along with PL, we developed a method to directly quantify spin density as a function of relative Raman and PL intensities alone, which is crucial for both fundamental research and practical applications.

\section{Results}
\textbf{Correlation of Raman and PL signals with spin defects.}
\begin{figure}[h]
    \centering
	\includegraphics[width=1.0\textwidth]{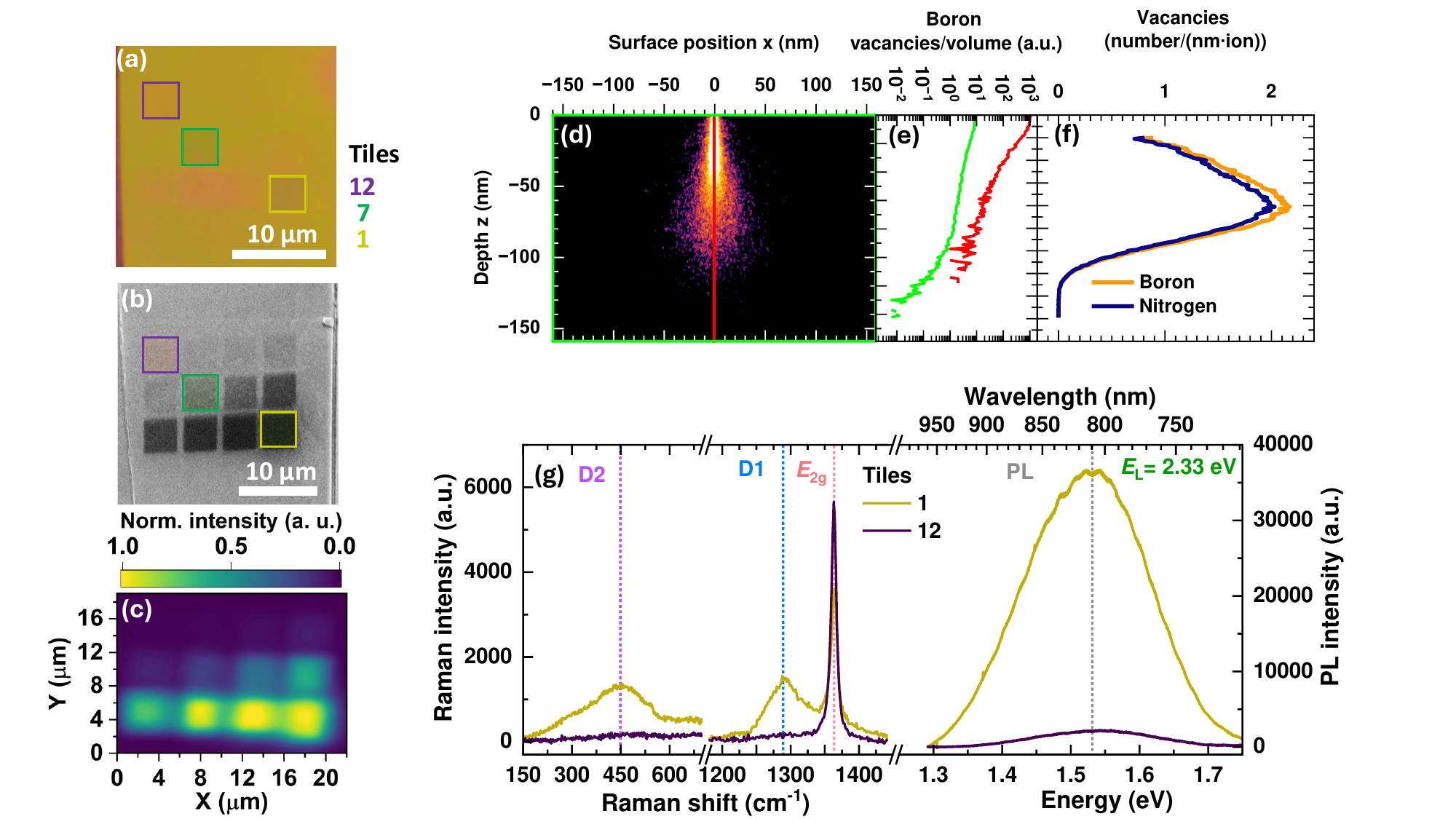}
	\caption{(a) Optical (b) SEM and (c) integrated PL images of irradiated hBN. The areas are labeled as Tiles 1-12, irradiated with decreasing ion fluence. Tile-1 (yellow) was irradiated with the highest fluence and Tile-12 (violet) with the lowest. For PL maps, spectra have been recorded point-by-point and the spectra have been integrated from \SI{1.37}{\electronvolt} to \SI{1.65}{\electronvolt}. (d) Simulated boron vacancy density created by a focused ion beam using the software SRIM. The depiction represents a cross section in the x-z plane where the x-y plane is parallel to the sample surface. The density is shown on a logarithmic color scale.
    (e)~Vertical line profiles for the red line shown in (d), while the green line is averaged over the x-dimension (green box around (d)). (f)~Depth distribution of the vacancy yield per ion distinguished by target atom. The data is the sum of every vacancy created at a certain depth regardless of the x or y position. (g) Characteristic spectra of the Raman modes D2, D1, \eg and PL for Tiles 1 and 12 with a laser excitation energy of $E_L=\SI{2.33}{\electronvolt}$ (532~nm).
    }
	\label{fig:optical-image}
\end{figure}
Bulk hBN was dry-transferred onto two substrates: SiO$_2$/Si (referred to as Sample-1) and a Au/Cu gold-coated copper stripline (referred to as Sample-2). \vb defects were introduced  by nitrogen irradiation at increasing fluences in a tile pattern (4 µm $\times$ 4 µm tiles). Details are provided in the Supporting Information (SI), refer to \autoref{tab:dose-Si} and  \autoref{tab:dose-Au}. 
Optical and scanning electron microscopy (SEM) images of Sample-1 are shown in  \autoref{fig:optical-image}  (a) and (b), respectively. The color contrast of the array of tiles displays the variation in fluence.
The PL map in {\autoref{fig:optical-image}} (c), obtained from integrated and normalized PL spectra, directly mirrors the irradiation pattern observed in the optical and SEM images.

The irradiation impact of the nitrogen focused ion beam was simulated using the software SRIM (The Stopping and Range of Ions in Matter) and is shown in \autoref{fig:optical-image} (d)-(f). 
For the calculations, monolayer collision steps and a full collision report were calculated for $\approx 10^4$ ions and analyzed afterwards (for more details see SI).
A lateral x-z cross section of the ion-induced damage plume is shown in (d) where the color of each pixel represents the amount of boron vacancies created per $\SI{1}{\nano\meter^3}$ on a logarithmic scale.
The x-y plane of the simulation is parallel to the sample surface and z is the depth below the surface. 
The profile along the red line in (d) is shown in sub panel (e). The green profile is additionally averaged over the x-axis.
When we take also the y-dimension into account, we get a comprehensive view of the vacancies created in a certain depth.
The data shown in panel (f) represent the vacancy yield per ion as a function of depth, obtained by dividing the average vacancies created at depth z by the number of ions used in the simulation. 
As can be seen, the defects are all created within $\approx$100~nm depth of the surface with approximately equal share of boron and nitrogen vacancies present. While this is a valid result, SRIM is not suitable to predict absolute defect densities as it treats damage cascades independently and does not consider secondary effects such as defect annealing, charge state, interstitials, etc.

\autoref{fig:optical-image} (g) shows the Raman and PL spectra measured at two tiles on Sample-2: Tile-1 (yellow) corresponds to the highest irradiation fluence, while Tile-12 (violet) represents the lowest. 
The whole spectrum was recorded in a single measurement with excitaion laser energy $E_\mathrm{L} = \SI{2.33}{\electronvolt}$ ($\lambda_\mathrm{L} = \SI{532}{\nano\meter}$). The Raman spectra are shown in \si{\per\centi\meter}, and the PL spectra in electronvolts (eV) for clarity. For Sample-1, the same signals are observed, although superimposed by Raman signals from the silicon substrate, as shown in \autoref{fig:Raman_Au_Si}. 

The most prominent mode of the Raman spectrum is at $\approx \SI{1365}{\per \centi\meter}$, which corresponds to the \eg phonon at the Brillouin zone center.
Notably, the peak position provides information about the nature of present isotopes due to the mass difference between $\mathrm{^{10}B}$ and $\mathrm{^{11}B}$ isotopes\cite{vuong2018isotope,janzen2024boron}. hBN with a single boron isotope retrieved from either $\mathrm{^{10}B}$-or $\mathrm{^{11}B}$-enriched hBN show the \eg mode at \SI[separate-uncertainty = true]{1394(1)} and \SI[separate-uncertainty = true]{1358(1)}{\per \centi\meter}, respectively. 
In pyrolytic BN (pBN), it appears at \SI[separate-uncertainty = true]{1367(1)}{\per \centi\meter} \cite{li2021defect}.
The next prominent features in the Raman spectrum are observed at $\approx \SI{1290}{\per \centi\meter}$  and \SI[separate-uncertainty = true]{450}{\per \centi\meter}, labeled as D1 and D2 respectively in \autoref{fig:optical-image} (g). 
These peaks are associated with \vb defects in hBN and pBN \cite{li2021defect,venturi2024selective,ren2023creation,linderalv2021vibrational}, and are notably absent for carbon-based defects in hBN\cite{mendelson2021identifying,koperski2020midgap}. 
The absence of transverse optical Raman modes associated with cubic-BN (cBN) at \SI[separate-uncertainty = true]{1055}{\per \centi\meter}  excludes the possibility of hBN-to cBN transition \cite{werninghaus1997raman,reich2005resonant}.

The characteristic and broad PL signature of \vb defects is observed at $\approx$ 1.53~eV (810~nm) at room temperature. In contrast to Raman modes, PL does not exhibit isotope dependence, as reported previously\cite{li2021defect,haykal2022decoherence,sasaki2023nitrogen}. 
\autoref{tab:Different modes} summarizes the Raman and PL peaks related to various defects in hBN, as observed in this study and as reported in the literature  \cite{li2021defect, venturi2024selective,ren2023creation, zhang2019improved,mendelson2021identifying,liu2022rational}. 
The peak position of the D1 mode in hBN is independent (non-dispersive) of the used laser energy $E_\mathrm{L}$, unlike the D mode in graphene\cite{ferrari2013raman}. We corroborate this finding for three different $E_\mathrm{L}$ as shown in  \autoref{fig:Raman_1800g}. The different dispersion relation in comparison to graphene is to be expected as we are not dealing with a semi-metal with a Dirac-cone band structure, but with an insulator.

\begin{table}[h]
	\centering
	\resizebox{\textwidth}{!}{ 
		\Large
		\begin{tabular}{c c c c c c}
			\rowcolor{lightgray}
			Type of defect & Process & \multicolumn{3}{c}{Raman modes (cm$^{-1}$)} & PL mode (eV) \\
			\rowcolor{lightgray}
			& & D2 & D1 & \eg &  \\\\
			 & nitrogen irradiation $^{\star}$  & 450 & 1290 & 1365 & 1.53 (810 nm)\\\\
			& neutron transmutation in $\mathrm{^{10}B}$ \cite{li2021defect} & 450 & 1296 & 1367.2 & 1.55 (800 nm)\\\\
			\vb & $^{12}\text{C}$, $^{20}\text{Ne}$, $^{69}\text{Ga}$ irradiation \cite{venturi2024selective}  &--  & 1290 & 1357 & 1.99 (620 nm) \& 1.53 (810 nm) \\\\
			& helium irradiation \cite{ren2023creation, carbone2025quantifying} & 448 & 1295 & 1367 & 1.5 (830 nm)\\\\
            \hline \\
			$\mathrm{V_BC_N^-}$ & carbon-doped \cite{zhang2019improved,mendelson2021identifying,liu2022rational} & -- & -- & 1366 & 2.11 (585 nm) \\
		\end{tabular}
		\caption{Summarizing the Raman and PL modes corresponding to different defects in hBN, as identified in the present study($\star$) and reported earlier\cite{li2021defect, venturi2024selective,ren2023creation, zhang2019improved,mendelson2021identifying,liu2022rational}.}
		\label{tab:Different modes}
	}
\end{table}

\begin{figure}[ht]
  \centering
  \includegraphics[width=1.0\textwidth]{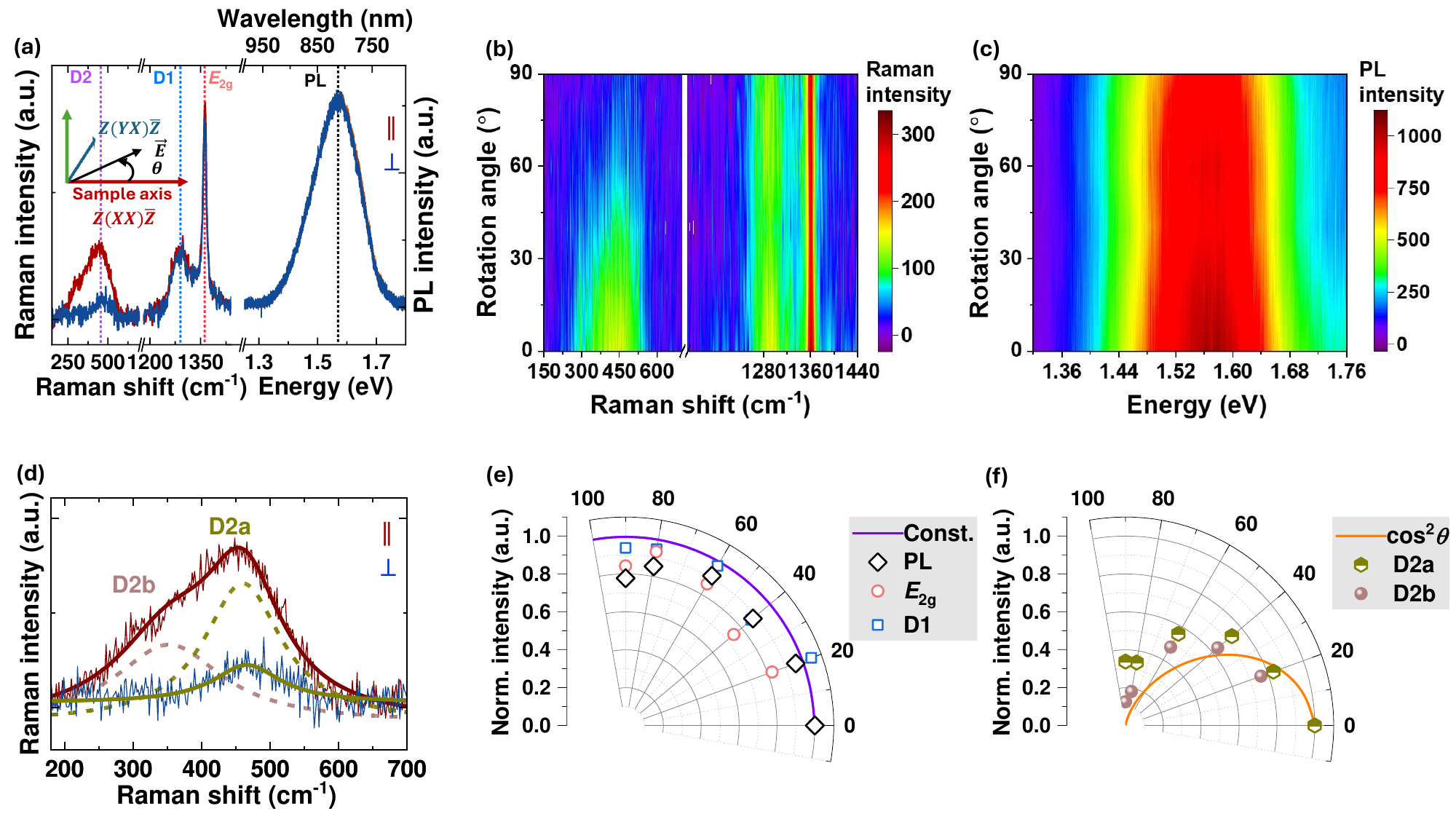}
      \caption{Linear polarization-dependent Raman modes: (a) Spectra corresponding to $Z(XX)\Bar{Z}$  and $Z(YX)\Bar{Z}$, are shown in  red and blue, respectively. The color contour plots present the variation of Raman (b) and PL signals (c) with respect to incident light polarization direction. (d) Lorentzian fitting of the D2 mode for two polarization configurations. For \textit{$\parallel$} polarization, the D2 peak is resolved into two sub-modes, D2a and D2b, appearing at $\approx$ 459	 and 352 cm$^{-1}$ with width $\Gamma \approx$ 138 and 191 cm$^{-1}$, respectively. In contrast, for $\perp$-polarization, D2b is not detected, and D2a appears with reduced intensity at 465  cm$^{-1}$ with $\Gamma$ $\approx$ 120 cm$^{-1}$.  (e,f) Polar plots of all Raman modes and the PL signal, normalized to their  $Z(XX)\Bar{Z}$ values. (e) Representing the isotropic behaviour of PL, \eg , and D$_1$. (f) $\cos^2{\theta}$ dependence of D2a and D2b.}
    \label{fig:Polarization}
\end{figure} 
\textbf{Polarization-dependence.}
To gain deeper insight into the origin of the Raman modes, we conducted linear polarization-dependent Raman measurements at Tile-4 (3.2 ions/nm$^2$) of Sample-2, as shown in \autoref{fig:Polarization}. 
 The polarization direction of the incident laser light is varied using a half wave plate ($\lambda/2$) which is introduced into the optical path of incident light, along with an analyzer in the scattered path. 
For a fixed incident laser polarization, the analyzer confines the scattered light either in horizontal or in vertical configuration.
\autoref{fig:Polarization} presents data for parallel polarization of excitation and emission. The orthogonal configuration exhibits similar behavior with reversed features as the $\lambda/2$ plate rotates relative to the sample axis with an angle $\theta$ (see \autoref{fig:vertical_analyzer} in the SI). 
By continuously varying the $\lambda/2$ plate, we can go from a $Z(XX)\Bar{Z}$ to $Z(YX)\Bar{Z}$ configuration, corresponding to parallel ($\parallel$) and perpendicular ($\perp$) polarization  measurements, respectively. This is illustrated in the inset of \autoref{fig:Polarization} (a).
In this notation, the first and fourth term represent the direction of incident and scattered light, respectively. While the second and third terms denote the polarization in the direction of the incident and scattered light, respectively. 
In \autoref{fig:Polarization} (a), spectra for \textit{ $\parallel$} and \textit{$\perp$} polarizations are shown in red and blue, respectively. 
In \autoref{fig:Polarization} (b) and (c), contour plots that capture the full polarization dependence of the Raman modes and PL are shown. 
These plots reveal that the intensities of the PL, \eg, and D1 modes remain constant against the rotation of the $\lambda/2$  plate. 
In contrast, the D2 mode demonstrates a pronounced dependence on it, highlighting its distinct behavior compared to the other modes.
A high-resolution measurement using a 1800 lines/mm grating reveals that the D2 mode for \textit{$\parallel$} polarization is comprised of two components, denoted as D2a and D2b.
In contrast, for $\perp$-polarization, only a single mode is observed, as presented in \autoref{fig:Polarization} (d). The intensities of different peaks from the contour plot in  \autoref{fig:Polarization} (b) and (c), normalized with respect to their $Z(XX)\Bar{Z}$ values, are depicted in \autoref{fig:Polarization} (e). 
This illustrates that only the D2 modes (D2a and D2b) follow the $\cos^2{\theta}$ dependence, as indicated by the solid orange line in \autoref{fig:Polarization} (f).  


The intensity of Raman modes is expressed as $I_\mathrm{R} \propto \left| e_{\text{in}}^T \cdot \hat{R} \cdot e_{\text{out}} \right|^2$, where $\hat{R}$  is the Raman tensor \cite{liu2017different}. 
The interaction between the in-plane polarization of the incident light and atomic vibrations alters the polarizability and the Raman tensor, which is reflected in the Raman signal.
Consequently, the intensity of the \eg mode exhibits an isotropic behaviour with respect to $\theta$, with an expected response of $I_R \propto$ constant. Such isotropic variation of \eg is reported for basal plane measurement in graphene \cite{cong2010raman} and transition metal dichalcogenide (TMDC) materials \cite{liu2021angle,yang2023angle}.

\begin{figure}[h]
	\centering
    \hspace*{-1.8cm}
	\includegraphics[width=1.2\textwidth]{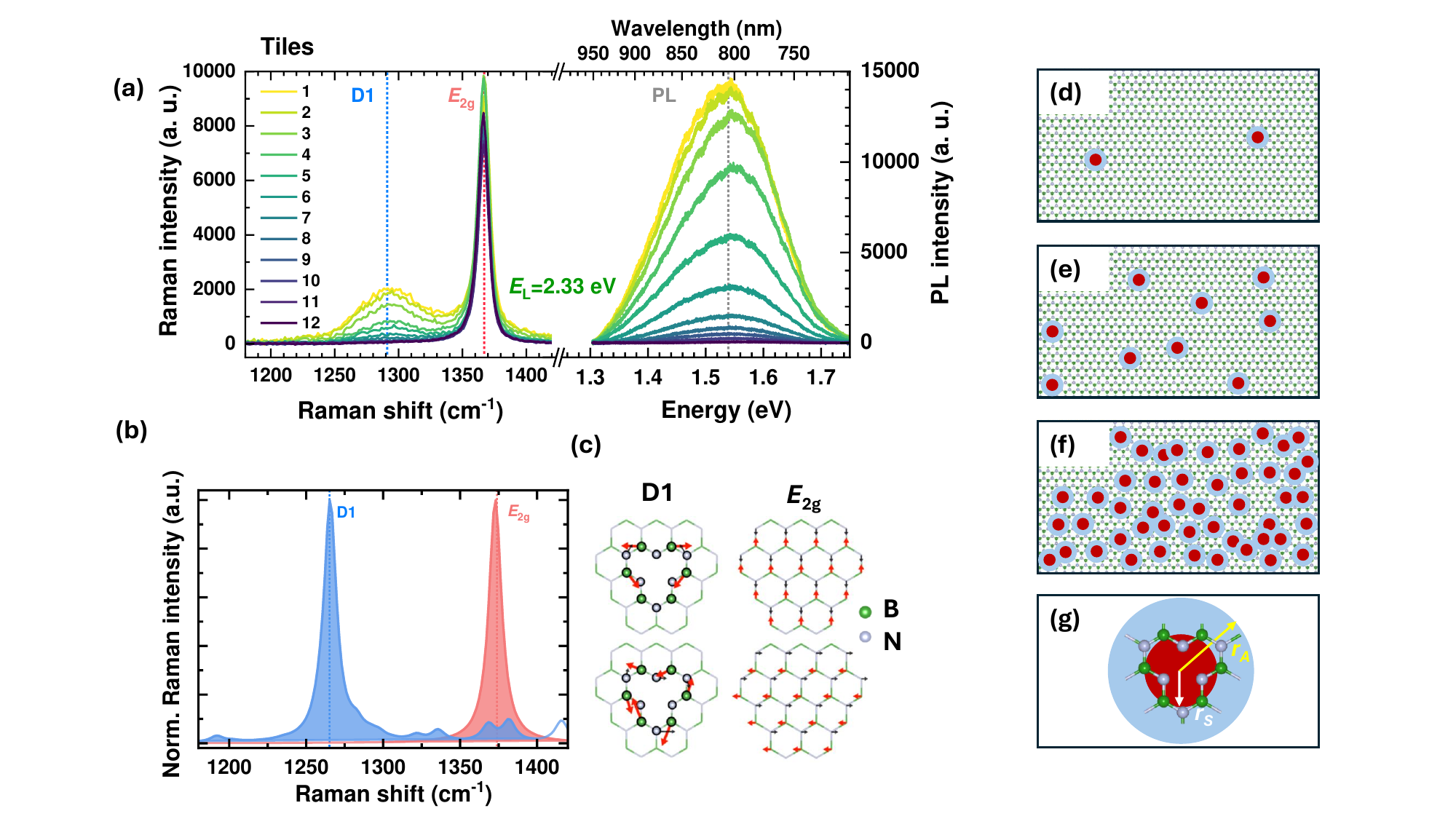}
	\caption{(a) Characteristic Raman and PL spectra recorded at different tiles, as shown in \autoref{fig:optical-image}  (a) and (b). (b) Calculated Raman spectrum and (c) Schematic diagram of atomic vibrations responsible for different Raman modes. (d)-(f) Schematic illustrations of defect densities for each irradiation fluence are presented in panels, adapted from \cite{lucchese2010quantifying}. As the defect density increases, the overlap becomes more pronounced, adversely affecting both the Raman and PL signals. (g) The red region represents the area directly impacted by ion collisions, while the light blue region indicates the extent to which the ion impacts remain visible.}
	\label{fig:Raman/PL}
\end{figure}

It is noteworthy that we are dealing with the \vb point defects with zero-field splitting parameter $D \approx \SI{3.5}{\giga \hertz}$ \cite{gottscholl2020initialization} . 
The positive value of $D$ describes that the electronic distribution around the defect center has an oblate shape. 
We believe that this oblate shape influences the electronic polarizability, which contributes to the observation of additional Raman modes. 
Further analysis of the Raman intensity with $\theta$ demonstrates a similar isotropic variation of \eg and D1 modes in \autoref{fig:Polarization} (e). 
This suggests that similar to the \eg-mode, in-plane atomic vibrations around the defect are associated with the D1 mode, as presented in schematic diagram \autoref{fig:Raman/PL} (c). 
However, for atomic vibrations out of the basal plane, the Raman intensity follows $I_\mathrm{R} \propto a \cos^2{\theta}$, as observed in TMDCs \cite{liu2021angle,hulman2019polarized}. 
A similar $\theta$ dependence is observed for both D2 modes in this study, as illustrated in \autoref{fig:Polarization} (f).

\newpage 

Furthermore, we studied the dependence on laser excitation energy ($E_\mathrm{L}$) of $\SI{2.62}{\electronvolt}$ (473~nm), $\SI{2.33}{\electronvolt}$ (532~nm), and $\SI{1.96}{\electronvolt}$ (633~nm) and find that PL and Raman spectra reveal a clear dependence as shown for Sample-2 in \autoref{fig:Raman/PL_wavelength_wavelength}. 
PL intensity peaks near the absorption maximum at $E_\mathrm{L}\approx$2.6~eV and remains strong even at 2.33~eV, which might be due to enhanced ionization of neutral boron vacancies ($\mathrm{V_B^0}$) into their optically active charged state \vb. \cite{gale2023manipulating} We find that \eg follows expected trends with excitation energy for a Raman mode. In contrast to that, the D1 mode mirrors the PL response, suggesting a shared origin linked to defect density and photo-induced ionization effects.

\textbf{Fluence-dependence.}  
\autoref{fig:Raman/PL} (a) shows the PL and Raman modes of Sample-1 at different tiles corresponding to varying irradiation fluences, as indicated in the images in \autoref{fig:optical-image}. 
An increase in irradiation fluence is expected to generate more spin defects. \cite{kianinia2020generation}
Consistent with this expectation, we observe an increase in intensity of the defect-related PL and D1 Raman modes with increasing irradiation fluence in \autoref{fig:Raman/PL} (a), while the intensity of \eg  remains almost constant within the experimental error. 
The observed similarity in the changes in D1 and PL signals indicates a possible direct correlation between the Raman and PL responses.

In addition to the observed intensity changes, the Raman line width $\Gamma$ of the \eg-mode provides information about the crystal structure. 
Specifically, the lower $\Gamma$  value for the \eg-mode at Tile-12, compared to Tile-1 shown in \autoref{fig:Raman width}, indicates that the crystallinity of hBN is better preserved for the low irradiation in Tile-12. 
To further illustrate this, the variations in $\Gamma$ over the irradiation fluence for the \eg and D1 modes, as well as for the PL width, are presented in \autoref{fig:Raman width}. 
These data sets highlight that Raman and PL width correlate with the irradiation fluence, reinforcing the impact of irradiation on the structural integrity of hBN.

To associate the Raman peaks in \autoref{fig:optical-image} (g) and \autoref{fig:Raman/PL} (a) with specific vibrational modes of the boron vacancy, we carried out DFT calculations using a supercell model with periodic boundary conditions. While the vibrational eigenmodes of defects in hBN, including the boron vacancy, have been previously studied with DFT\cite{linderalv2021vibrational}, the mere presence of such modes does not guarantee their appearance in the Raman spectrum. Here, to calculate the Raman activities of the vibrational modes of \vb, we considered a non-resonant Raman process, assuming an average over the polarizations of incident and scattered light\cite{lazzeri2003ramandft}. The Raman scattering efficiency was approximated by non-spin-polarized calculations, while ensuring that the calculated vibrational modes are consistent with those of the triplet state. As shown in red in \autoref{fig:Raman/PL} (b), the frequency of the $E_\mathrm{2g}$ mode obtained from our calculations for a pristine hBN monolayer (8×8 unit cells, 128 atoms) aligns well with the experiment. When \vb is introduced into the supercell, the calculated Raman spectra reveal prominent defect-induced Raman-active modes in the frequency regions corresponding to the experimental D1 and D2 peaks (\autoref{fig:Raman/PL} (b) and \autoref{fig:DFT_8x8layer_full}).

The double-degenerate vibrational mode at 1265 \si{\per \centi \meter} (at the D1 peak) exhibits a large amplitude and is strongly localized near the defect, identifying it as a typical “local mode”\cite{maradudin1965defectvibrations}. Its vibrational pattern is shown in \autoref{fig:Raman/PL} (c). The calculated inverse participation ratio (IPR), used to quantify the degree of localization, is 11.6, indicating that this mode is predominantly localized on only approximately 12 atoms in the supercell\cite{alkauskas2014nvcenters}. This explains why its frequency is independent of the supercell size (cf. \autoref{fig:DFT_size}) and is largely resilient to variations in the local defect concentration within the sample.

\textbf{Phenomenological model.}
In the determination of defect density, refs. \cite{lucchese2010quantifying,canccado2011quantifying} emphasize  the importance of considering the distance between point like defects.
These studies propose an empirical model based on the intensity ratio of D and \eg Raman modes in graphene.
 \autoref{fig:Raman/PL} (d) - (f) show schematics of the crystal structure containing defects, adapted from ref \cite{lucchese2010quantifying}. 
The phenomenological model is derived by solving geometrical rate equations considering the evolution of a structurally deformed region (S-, shown in red in \autoref{fig:Raman/PL} (g)) expanding up to a radius $r_\mathrm{S}$ and adjacent activated region (A-, shown in light blue in \autoref{fig:Raman/PL} (g)) expanding up to $r_\mathrm{A}$ with irradiation fluence. 
Beyond $r_\mathrm{S}$, the lattice structure is preserved. 
Nevertheless, the proximity to defects leads to a mixing of Bloch states near high-symmetry points up to distance ${r_\mathrm{A}}$, breaking selection rules and resulting in the emergence of D-Raman modes.
The derived equation for the ratio of the Raman intensities is given by \cite{lucchese2010quantifying}:
\begin{equation}\label{eq:eqn-1}
	\frac{I_\mathrm{D}}{I_\mathrm{\textit{E}_{2g}}} = C_\mathrm{A} \frac{r_\mathrm{A}^2 - r_\mathrm{S}^2}{ r_\mathrm{A}^2 - 2r_\mathrm{S}^2 } \left[\exp\left(-\frac{\pi r_\mathrm{S}^2}{L_\mathrm{D}^2}\right) - \exp\left(-\frac{\pi \left( r_\mathrm{A}^2 - r_\mathrm{S}^2 \right)}{L_\mathrm{D}^2}\right)\right] + C_\mathrm{S} \left[1 - \exp\left(-\frac{\pi r_\mathrm{S}^2}{L_\mathrm{D}^2}\right)\right] 
\end{equation}

The distance between defects is considered by the characteristic length $L_\mathrm{D}$. $C_\mathrm{A}$ and $C_\mathrm{S}$ are scaling factors for activation and structural deformation, respectively.

\begin{figure}[t]
  \centering
  \includegraphics[width=0.99\textwidth]{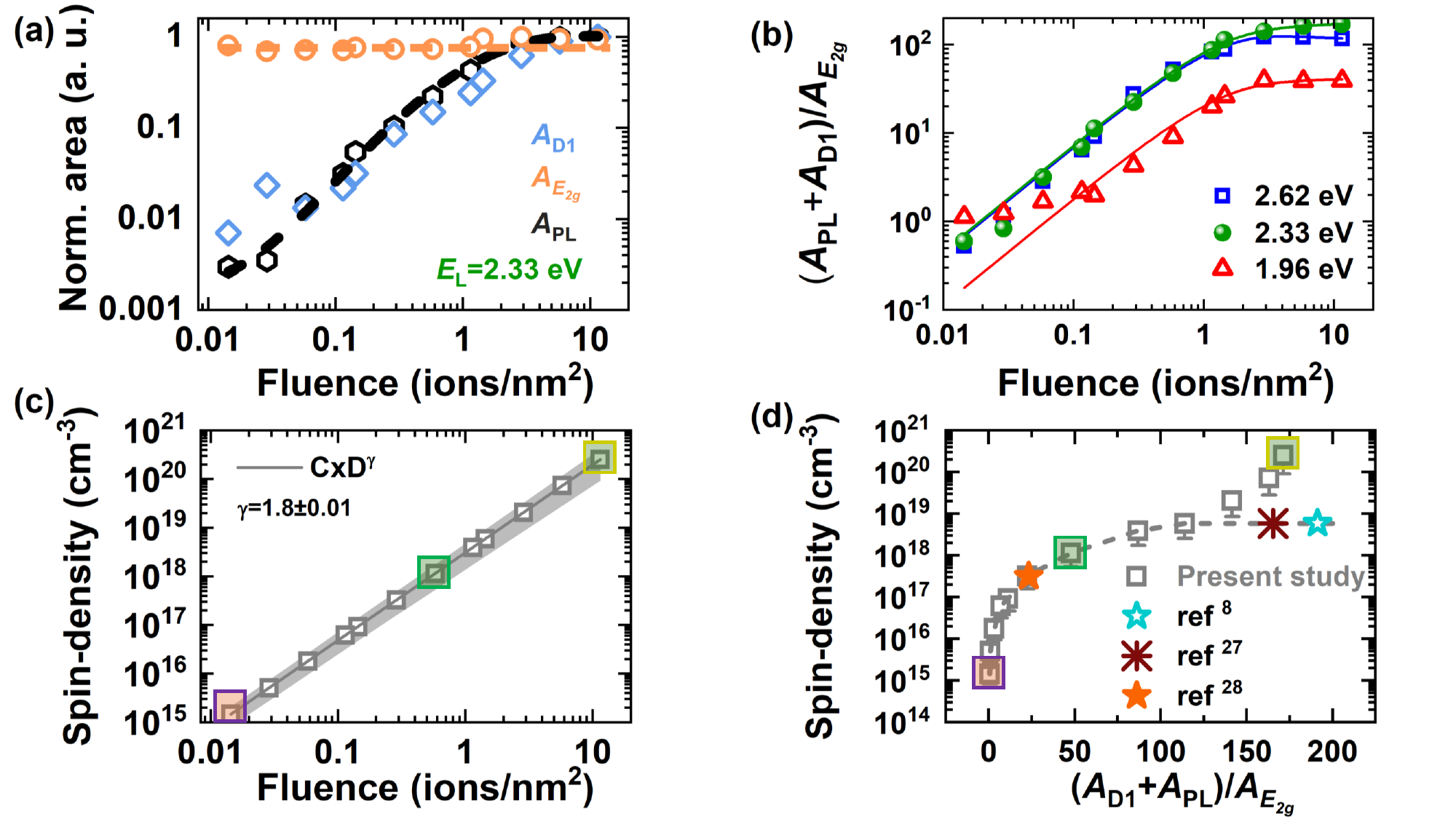}
      \caption{(a) Dependence of integrated area $A$ under curve for D1, \eg Raman modes and PL at $E_\mathrm{L} = \SI{2.33}{\electronvolt}$ on the irradiation fluence. Dotted traces are guides to the eye. 
      (b) Data points represent $(A_\mathrm{D1}+A_\mathrm{PL})/A_{E_\mathrm{2g}}$, and the solid traces are calculated following \autoref{eq:eqn-2} for three $E_\mathrm{L}$ values. 
      (c) The calculated spin density using \autoref{eq:eqn-2}.
      The shaded area represents the error associated with the fitting. 
      The solid line is a fit of  $ C\cdot D^{\gamma}$ with $\gamma=1.8\pm0.01$, representing the power law dependence of the spin-defect density on the irradiation fluence. The colored boxes mark the values for the corresponding tiles in Fig. 1.
      (d) The plot illustrates the direct estimation of spin defect density from the ratio $(A_\mathrm{D1}+A_\mathrm{PL})/A_{E_\mathrm{2g}}$ in uncharacterized systems based on spectral response. The dashed line is a guide to the eye. The values  $(A_\mathrm{D1}+A_\mathrm{PL})/A_{E_\mathrm{2g}}$ $\approx 191$ \cite{gottscholl2021spin},$\approx 165$ \cite{venturi2024selective} and $\approx 20$ \cite{ren2023creation}, calculated from spectral response data in these references, following the described method are marked by different asterisks.
      	}
    \label{fig:Fitting}
\end{figure}

We fitted all Raman and PL spectra and used the areas for further analysis as shown in \autoref{fig:Raman/PL_wavelength_wavelength}.
In our further analysis, we exclude the delocalized D2 mode due to its spectral overlap with silicon modes and its weak intensity.
The areas of D1 ($A_\mathrm{D1}$), $\mathrm{\textit{E}_{2g}}$ (A$_\mathrm{\textit{E}_{2g}}$), and PL ($A_\mathrm{PL}$) are are plotted as a function of irradiation fluence in  \autoref{fig:intensity_EL} for all three $E_\mathrm{L}$. 
Notably, the intensity of $A_\mathrm{PL}$ is approximately two orders of magnitude higher than $A_\mathrm{D1}$. 
However, when normalizing these variations relative to their maximum values, as shown in  \autoref{fig:Fitting} (a), a correlation between $A_\mathrm{PL}$ and $A_\mathrm{D1}$ becomes apparent --
i.e. the relative responses of $A_\mathrm{PL}$ and $A_\mathrm{D1}$ to ion fluence changes are similar. 
%
We therefore introduce a single metric ($\frac{A_\mathrm{D1}}{A_{\textit{E}_\mathrm{2g}}}$+$\frac{A_\mathrm{PL}}{A_\mathrm{\textit{E}_{2g}}}$) to make use of both fluence-dependent \vb signatures normalized to the \eg reference. 
Still, using only $\frac{A_\mathrm{D1}}{A_{\textit{E}_\mathrm{2g}}}$ or $\frac{A_\mathrm{PL}}{A_\mathrm{\textit{E}_{2g}}}$ is sufficient for this analysis -- however with reduced SNR.
While maintaining the same physical meaning of $r_\mathrm{S}$, $r_\mathrm{A}$ and $L_\mathrm{D}$, we adjust the scaling factors for activation  ($C_\mathrm{A}^{\prime}$) and structural deformation ($C_\mathrm{S}^{\prime}$) to apply to both $A_\mathrm{D}$ and $A_\mathrm{PL}$. 
Consequently, \autoref{eq:eqn-1} is modified to:
\begin{equation}\label{eq:eqn-2}
	\frac{A_\mathrm{D1}+A_\mathrm{PL}}{A_\mathrm{\textit{E}_{2g}}} = C_A^{\prime} \frac{r_\mathrm{A}^2 - r_\mathrm{S}^2}{ r_\mathrm{A}^2 - 2r_\mathrm{S}^2 } \left[\exp\left(-\frac{\pi r_\mathrm{S}^2}{L_\mathrm{D}^2}\right) - \exp\left(-\frac{\pi \left( r_\mathrm{A}^2 - r_\mathrm{S}^2 \right)}{L_\mathrm{D}^2}\right)\right] + C_\mathrm{S}^{\prime} \left[1 - \exp\left(-\frac{\pi r_\mathrm{S}^2}{L_\mathrm{D}^2}\right)\right]
\end{equation}

The PL is solely associated with the spin carrying \vb defect, as follows from optically detected magnetic resonance \cite{gottscholl2020initialization, haykal2022decoherence}. According to our previous discussion, it follows that D1 can also be associated with the spin defect. Hence, we assume that the characteristic length LD that we extract represents the spacing between spin-defect.
We can further relate $L_\mathrm{D}$ to the irradiation fluence through $L_\mathrm{D}=\frac{\alpha}{\text{Fluence}^\beta}$, where $\alpha$ is a proportionality factor and $\beta$ is a fluence-dependence exponent. We then fit \autoref{eq:eqn-2} to the data in \autoref{fig:Fitting} (b) for all three $E_\mathrm{L}$. $r_\mathrm{A}$, ${\alpha}$, and ${\beta}$ are set as global parameters shared across all three $E_\mathrm{L}$, while $C_\mathrm{A}^{\prime}$ and $C_\mathrm{S}^{\prime}$ are specific to each individual $E_\mathrm{L}$, as shown in \autoref{fig:variation with EL}. 
The fitting yields ${\alpha} = 4.27 \pm 0.93$, ${\beta} = 0.6 \pm 0.02$, and $r_\mathrm{A} = 2.42 \pm 0.43 $ nm, with $r_\mathrm{S}$ = 1.0 nm (as determined for structural disorder \cite{lucchese2010quantifying} ). 
In graphene the Raman relaxation length is reported to be $\approx$ 2--4 nm \cite{lucchese2010quantifying,canccado2011quantifying}. 
We find $r_\mathrm{A}$-$r_\mathrm{S}$ $\approx$ 1.4 nm, which suggests a shorter Raman relaxation length of the D1 mode in hBN. 
Following from this analysis, we now calculate the volume spin density, $n_\mathrm{D}=\frac{10^{21}}{(4/3)\pi L_\mathrm{D}^3}$ in \si{\centi \meter^{-3}}, as illustrated in \autoref{fig:Fitting} (c), with the shaded area depicting the potential error associated with the fitting. 
Furthermore, the estimated spin defect density adheres a power law with respect to irradiation fluence $C\cdot D^{\gamma}$, as shown by the solid line in the figure, with an exponent of $\gamma=1.8\pm0.01$. From this graph, the spin density can be read out directly as function of 30~keV nitrogen irradiation fluence.

Finally, to enable a straightforward estimation of spin-defect density by a single spectral measurement $independent$ of irradiation type, we plot the spin-defect density against the ratio $(A_\mathrm{D1}+A_\mathrm{PL})/A_\mathrm{\textit{E}_{2g}}$ in \autoref{fig:Fitting} (d). 
We observe that for a range of four orders of magnitude up to a density of 10$^{19}$ \SI{}{\per\centi\meter\cubed} the model accurately describes the dependence on the extracted Raman and PL intensities.
Therefore, by measuring Raman and PL signatures of \vb, with this graph alone, it is possible to determine the spin defect density. 
If experimentally either D1 or PL are not accessible, this analysis is still valid by using only the ratio of A$_\mathrm{D1}$/$A_{E_\mathrm{2g}}$ or A$_\mathrm{PL}$/$A_{E_\mathrm{2g}}$ as shown in \autoref{fig:calculation_Raman_PL}. Our measurements further exhibit that variations in experimental conditions, such as objective magnification and grating grooves, only affect $(A_\mathrm{D1}+A_\mathrm{PL})/A_\mathrm{\textit{E}_{2g}}$  by less than 15 $\%$ (see \autoref{fig:expt condition}).


Limits of the model are discussed in the following.
In \autoref{fig:Fitting} and \autoref{fig:intensity_EL}, we observe saturation or a decrease in PL intensity at high fluence above \SI{1}{ion\per\nano\meter\squared}, indicating damage to the crystal structure leading to deteriorating influences on the \vb system. 
The crystal damage is also reflected in the broadening of the \eg mode and PL width in \autoref{fig:Raman width}. 
This indicates that the model is not suited for 30~keV nitrogen ion fluences higher than \SI{1}{ion\per\nano\meter\squared} (=\SI{1E14}{ions\per\cm\squared}) and imposes a limit. However, this is specific for this type of irradiation. More general, the model is accurate until the onset of PL intensity saturation and \eg broadening for any type of irradiation.
It has to be noted that this approach extracts the defect density implicitly from \autoref{eq:eqn-2}. 
This introduces some ambiguity when inverting the formula and becomes relevant for high density as seen for the last points in \autoref{fig:Fitting} (d), which are therefore expected to deviate. 

For comparison, we calculate  $(A_\mathrm{D1}+A_\mathrm{PL})/A_{\textit{E}_\mathrm{2g}}$ from the spectral data given in ref. \cite{gottscholl2021spin} (see \autoref{fig:Raman_PL_ref}). 
The estimated spin density of $\approx 5.9 \times 10^{18}$ cm$^{-3}$, obtained using the present method, is indicated by an asterisk in \autoref{fig:Fitting} (d). 
The reported defect density from EPR measurements is $\approx 5.4 \times 10^{17}$ cm$^{-3}$, showing a one-order-of-magnitude difference. 
We ascribe this discrepancy to differences in defect creation processes.
Neutron irradiation is highly boron 10-isotope selective \cite{li2021defect} and the irradiation fluence that was used for the sample is considered high. This implies that the irradiation is in a region beyond the limits of our model.
Furthermore, we expect neutron irradiation to create primarily boron vacancies over other defects \cite{Knoll.2010}. 
This selectiveness might have severe influence on the Raman active modes. 
The 3D nature of the neutron irradiated crystal does also impact the \eg mode that we used for our model. 
To accurately apply our model to neutron irradiated samples, a similar irradiation study to this work would have to be repeated. This would also enable comparison of our model to the estimation of defect density with EPR.
We expect our model to be much more comparable with other types of ion irradiation.

Based on our findigs, we calculate $(A_\mathrm{D1} + A_\mathrm{PL})/A_{\textit{E}_\mathrm{2g}}$ from the reported spectra, finding values of approximately 165 (Ga-irradiated at $1 \times 10^{15}$ ions/cm$^2$) \cite{venturi2024selective} and 20 (He-irradiated at $2 \times 10^{14}$ ions/cm$^2$) \cite{ren2023creation}, which correspond to spin defect densities of around $6.2 \times 10^{18}$ cm$^{-3}$ and $3.2 \times 10^{17}$ cm$^{-3}$, respectively. These values are also indicated by asterisks in \autoref{fig:Fitting} (d).

\section{Conclusion} 
In conclusion, our results demonstrate that the new D1 and D2 Raman modes serve as signature peaks of \vb defects in hBN. 
Using polarized Raman spectroscopy and DFT calculations, we investigated the origin of these modes and their relation to  atomic vibrational eigenmodes of defects in hBN.
We observed that both the intensity of PL and defect-related Raman modes directly depend on the irradiation fluence.
Based on these observations, we established an empirical relation linking the integrated area of the D1, \eg and PL modes to the irradiation fluence. The derived model is universally applicable independent of irradiation type and the only method available for thin hBN flakes.
This allows for an easy-to-use all-optical quantification of absolute spin defect density present in hBN, making it an indispensable tool for photonic applications of spin defects in hBN.




\newpage
\section{Method}

\textbf{Experimental.}
  Raman and PL spectra were measured in backscattering configuration using a commercial micro-Raman system (Horiba LabRAM HR800) at room temperature with a 600 lines/mm grating (spectral resolution $\approx \SI{3.00}{\per\centi\meter}$ per CCD pixel). 
  A 100x objective (Olympus, NA=0.9) with laser spot size of $\approx \SI{1}{\micro \meter}$ was used, and the laser power was consistently maintained at $\approx \SI{1}{\milli \watt}$. 
  
\textbf{Computational Methods.}
All DFT calculations were performed using the Quantum ESPRESSO package v.7.3\cite{giannozzi2009quantum, giannozzi2017advanced}, employing a plane-wave basis set and norm-conserving pseudopotentials. For the \vb defect, we considered different supercell sizes with one missing boron atom each, including a single hBN monolayer (4×4 unit cells / 31 atoms, 6×6 unit cells / 71 atoms, and 8×8 unit cells / 127 atoms) and bulk hBN (8×8×2 unit cells / 511 atoms with four monolayers). The convergence threshold for atomic position relaxation was set to $10^{-5}$ Ry/Bohr. To avoid spurious distortions in defect-containing supercells, particularly for smaller systems, all calculations were conducted using the equilibrium lattice constants of pristine hBN. However, we note that fixing the lattice constants in some cases can induce spurious imaginary frequencies in the vibrational spectrum. We find the calculated electronic structure and vibrational modes to be robust with respect to the use of either the PBE exchange-correlation functional\cite{perdew1996gga} within the generalized gradient approximation (GGA) or the local density approximation (LDA). For computational reasons, we use the LDA for the calculation of the Raman coefficients.

The vibrational modes of pristine and defect-containing supercells were computed using density functional perturbation theory (DFPT)\cite{baroni2001phononsdft} as implemented in the PHONON module of Quantum ESPRESSO. The localization of each vibrational mode k was quantified using the inverse participation ratio (IPR)\cite{bell1970localization, alkauskas2014nvcenters}, defined as:

\begin{equation}\label{eq:si_ipr}
\text{IPR}_k = \frac{1}{\sum_a \left( \sum_i \Delta r_{ai,k}^2 \right)^2},
\end{equation}

where $\Delta r_{ai,k}$ is the normalized displacement vector of atom $a$ along direction $i$ in phonon mode $k$. Nonresonant Raman coefficients were then calculated based on DFPT results by evaluating the second-order response to a uniform electric field\cite{lazzeri2003ramandft}. The computed Raman activities were convoluted with a Lorentzian lineshape of 10 \si{\per \centi \meter} width.

\begin{acknowledgement}


V.D. acknowledges the support by the European Research Council (ERC) (Grant agreement No. 101055454). A.P., P.K., S.H. and V.D. acknowledges the funding by the lighthouse project IQ-Sense of the Bavarian State Ministry of Science and the Arts as part of the Bavarian Quantum Initiative Munich Quantum Valley (15 02 TG 86). S.H., A.S., V.D. acknowledge financial support from the Würzburg-Dresden Cluster of Excellence on Complexity and Topology in Quantum Matter ct.qmat (EXC 2147, DFG project ID 390858490). 

L.S., T.T. and I.A. acknowledge the Australian Research Council (CE200100010) and the Asian Office of Aerospace Research and Development (FA2386-20-1-4014) for financial support.

T.B. and W.G.S. acknowledge computational resources provided by the Paderborn Center for Parallel Computing (PC$^2$).

We would also like to thank the authors of refs.\cite{ren2023creation,venturi2024selective} for the valuable discussions and for providing spectral data used in comparison with our analysis.

\end{acknowledgement}
\bibliography{Draft_4}

\newpage
\renewcommand{\thefigure}{S\arabic{figure}}
\renewcommand{\thetable}{S\arabic{table}}
\setcounter{figure}{0}
\setcounter{table}{0}
\setcounter{equation}{0} 
\setcounter{enumi}{0} 
\setcounter{enumiv}{0} 
\setcounter{page}{1}

\begin{center}
\textbf{\large{Supporting Information}} \\
\mciteErrorOnUnknownfalse
\textbf{Quantifying Spin Defect Density in hBN via Raman and Photoluminescence Analysis} \\
Atanu Patra,$^\dag$ Paul Konrad,$^\ddagger$ Andreas Sperlich,$^{\ddagger,\perp}$ Timur Biktagirov,$^\P$ Wolf Gero Schmidt,$^\P$ Lesley Spencer,$^\S$ Igor Aharonovich,$^{\S,\parallel}$ Sven Höfling,$^{\dag,\perp}$ and Vladimir Dyakonov$^{\ddagger,\perp}$

\textit{$^\dag$Julius-Maximilians-Universität Würzburg, Lehrstuhl für Technische Physik, Am Hubland, Würzburg 97074, Deutschland\\
$^\ddagger$Julius-Maximilians-Universität Würzburg, Experimental Physics 6, Am Hubland,
Würzburg 97074, Deutschland\\
$^\P$Universität Paderborn, Department Physik, Warburger Str. 100, 33098 Paderborn, Deutschland\\
$^\S$School of Mathematical and Physical Sciences, University of Technology Sydney, Ultimo, New South Wales 2007, Australia\\
$^\perp$Physikalisches Institut and Würzburg-Dresden Cluster of Excellence ct.qmat, Deutschland\\
$^\parallel$ARC Centre of Excellence for Transformative Meta-Optical Systems, University of Technology Sydney, Ultimo, New South Wales 2007, Australia}
\end{center}

\newpage
\textbf{Details of nitrogen irradiation fluence.}
The specifics of the nitrogen irradiation fluence are outlined in \autoref{tab:dose-Si} and \autoref{tab:dose-Au} for Sample-1 (SiO$_2$/Si) and Sample-2 (gold-coated copper substrates, Au/Cu), respectively. 

\begin{table}[H]
	\centering
	\resizebox{0.7\textwidth}{!}{ 
		\begin{tabular}{c c c }
			\rowcolor{lightgray}
			Tile number & nitrogen irradiation   & nitrogen irradiation  \\
			\rowcolor{lightgray}
			& $@$ $10^{11}$ ions/cm$^2$ & ions/nm$^2$   \\\\
			1	& 11500 & 11.5   \\
			2	& 5800 & 5.8   \\
			3	& 2890 & 2.89   \\
			4	& 1440 & 1.44   \\
			5	& 1150 & 1.15   \\
			6	& 580 & 0.58   \\
			7	& 289 & 0.289   \\
			8	& 144 & 0.144   \\
			9	& 115 & 0.115   \\
			10	& 58 & 0.058   \\
			11	& 28.9 & 0.0289   \\
			12	& 14.4 & 0.0144   \\
		\end{tabular}
	}
	\caption{ A 30 keV nitrogen  irradiation fluence was applied for different tiles on a hBN flake $after$ dry-transfer onto the SiO$_2$/Si substrate.}
	\label{tab:dose-Si}
\end{table}

\begin{table}[H]
	\centering
	\resizebox{0.7\textwidth}{!}{ 
		\begin{tabular}{c c c }
			\rowcolor{lightgray}
			Tile number & nitrogen irradiation  & nitrogen irradiation  \\
			\rowcolor{lightgray}
			&  $@$ 3.2$\times$ $10^{11}$ ions/cm$^2$ & ions/nm$^2$   \\\\
			1	& 7000 & 22.4   \\
			2	& 4000 & 12.8   \\
			3	& 2000 & 6.4   \\
			4	& 1000 & 3.2   \\
			5	& 700 & 2.2   \\
			6	& 400 & 1.3   \\
			7	& 200 & 0.6   \\
			8	& 100 & 0.3   \\
			9	& 70 & 0.2   \\
			10	& 40 & 0.13   \\
			11	& 20 & 0.06   \\
			12	& 10 & 0.03   \\
		\end{tabular}
	}
	\caption{A similar 30 keV nitrogen irradiation was performed on hBN $before$ dry-transfer onto the gold-coated copper substrate.}
	\label{tab:dose-Au}
\end{table}

\textbf{Simulation details.}
For the simulation of irradiation damage, the software SRIM (The Stopping and Range of Ions in Matter) was used. 
The target material was a 2000\si{\angstrom} thick boron-nitride compound with 50-50 atomic percent (56.4 to 43.5 mass percent) for nitrogen and boron, respectively. A pre-calculated density of \SI{2.1}{\g \per \centi\meter} was used. 
The simulation displacement energy ${E_\mathrm{D}}$, binding energy ${E_\mathrm{B}}$, and surface energy ${E_\mathrm{S}}$ were set to ${E_\mathrm{D}=\SI{20,96}{\electronvolt}}$, $E_\mathrm{B}=\SI{3}{\electronvolt}$, $E_\mathrm{S}=\SI{20,96}{\electronvolt}$ for nitrogen and
$E_\mathrm{D}=\SI{17,6}{\electronvolt}$, $E_\mathrm{B}=\SI{3}{\electronvolt}$, $E_\mathrm{S}=\SI{17,6}{\electronvolt}$ for boron ~${}^1$. 
The simulation yields a full collision report for ions and recoils. The file was analyzed for collisions that resulted in the formation of a vacancy.

\begin{figure}[H]
	\centering
	\includegraphics[width=0.95\textwidth]{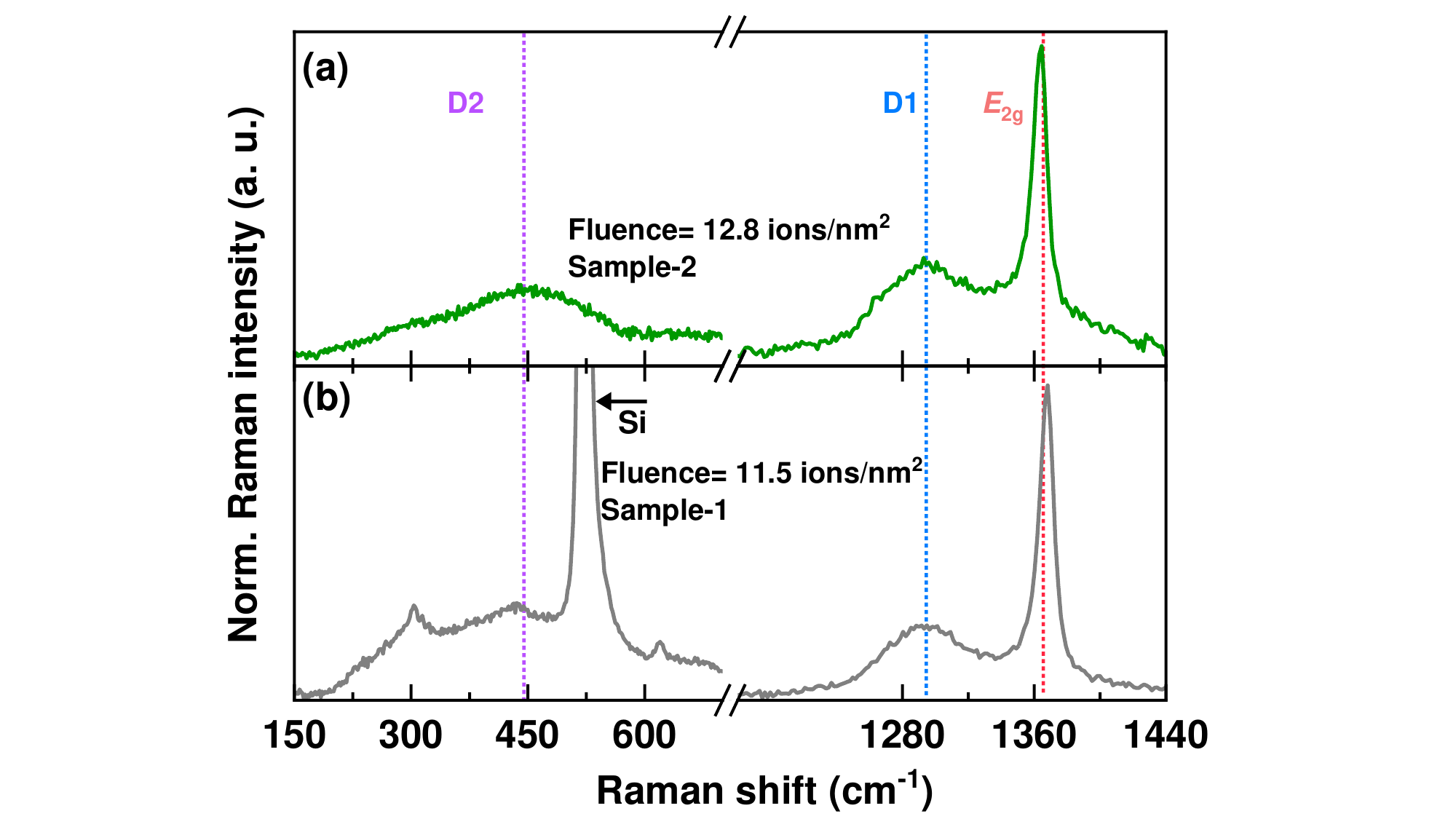}
	\caption{Raman spectra of irradiatd hBN on Au/Cu (a) and SiO$_2$/Si substrates (b). }
	\label{fig:Raman_Au_Si}
\end{figure}

\textbf{Different substrates.}
\autoref{fig:Raman_Au_Si} demonstrates the Raman modes of irradiated hBN, with (a) showing results for the Au/Cu substrate (Sample-2) and (b) for the SiO$_2$/Si substrate (Sample-1). The Raman spectra exhibit three prominent peaks: \eg, D1, and D2. Although the strong silicon peak overlaps with the D2 mode, the D1 peak is still observable in \autoref{fig:Raman_Au_Si}~(b).

\newpage 

\begin{figure}[H]
	\centering
	\includegraphics[width=0.95\textwidth]{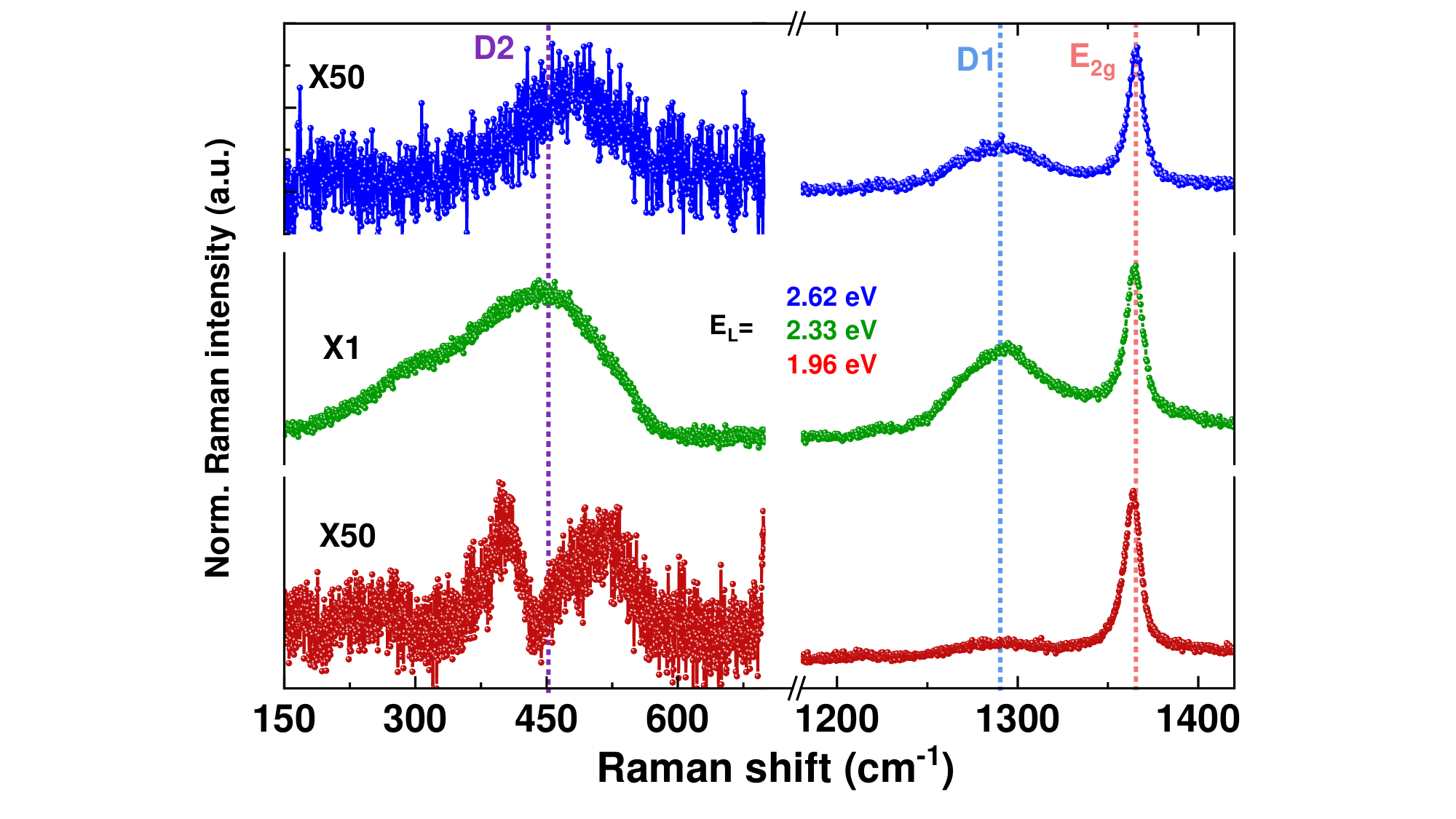}
	\caption{Raman spectra recorded using a 1800 lines/mm grating at three different laser excitation energies, $E_L$. The independence of the D1 peak with the excitation energy exhibits a non-dispersive nature unlike graphene. To enhance the visibility of the D2 peak, the spectra in the range of 150 to 700 cm$^{-1}$  are multiplied by a factor of 50 for the 2.62 eV and 1.96 eV excitation energies.}
	\label{fig:Raman_1800g}
\end{figure}

\textbf{High resolution Raman measurement.} 
The Raman spectrum shown in \autoref{fig:Raman_1800g} is measured with a grating of 1800 lines/mm at Tile-4, nitrogen fluence of 3.2 ions/nm$^2$ of Sample-2. We find that the D1 Raman mode remains unaffected by the change in excitation energy. Conversely, the cause behind the observed dependence in the D2 mode is still uncertain and necessitates further analysis.

\newpage

\begin{figure}[H]
	\centering
    \includegraphics[width=   0.95\textwidth]{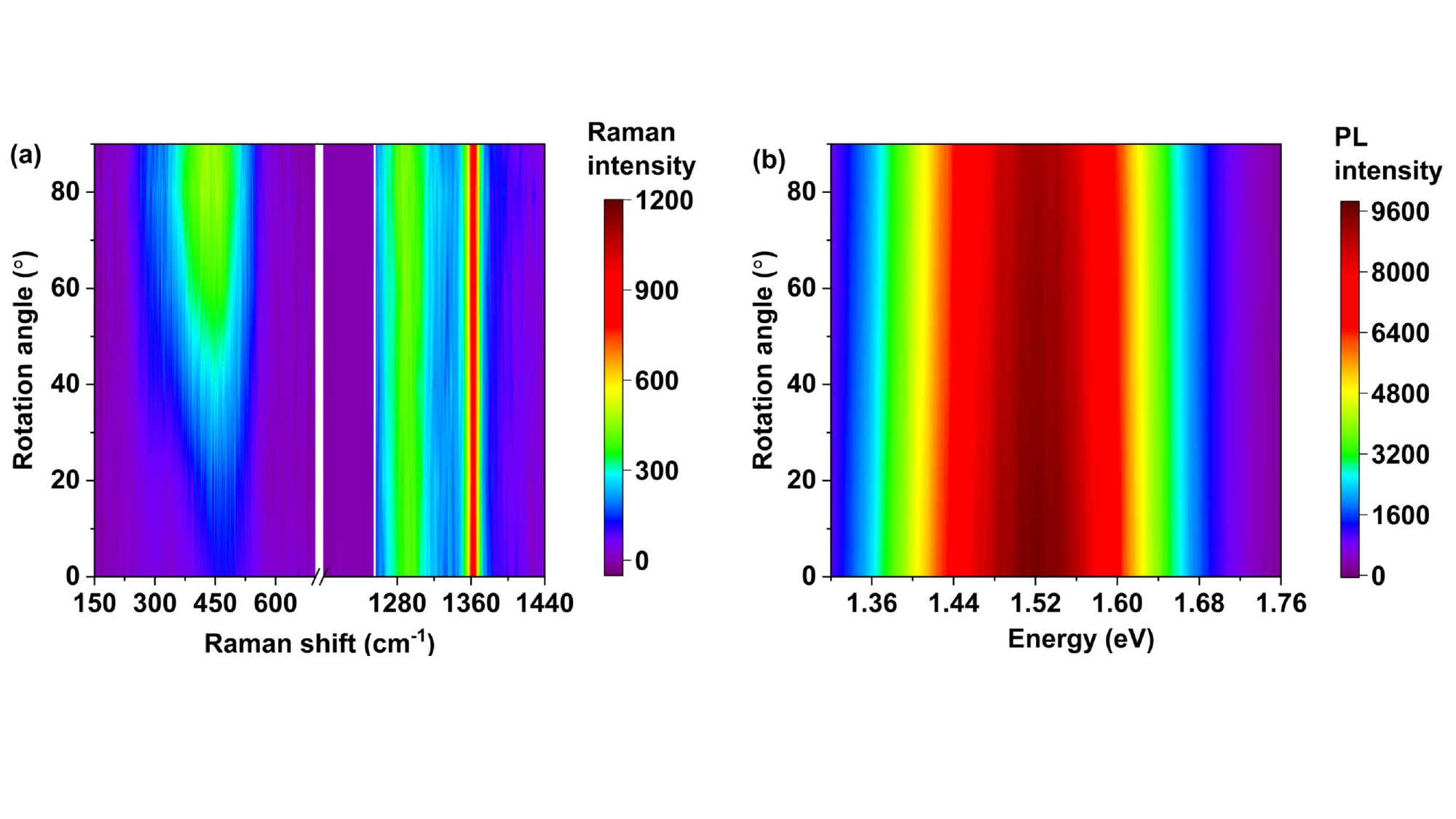}
	\vspace{-1.5cm}
    \caption{Variation of Raman modes (a) and photoluminescence (PL)  signal (b) with a perpendicular analyzer.  As shown in the main text of \autoref{fig:Polarization} (b) and (c), here the pattern is reversed. Notably only D2 mode shows a polarization dependence.}
	\label{fig:vertical_analyzer}
\end{figure}

\textbf{Polarization-dependence.}
The contour color plot displays the D2, D1, \eg , and PL modes as a function of the incident polarization in \autoref{fig:vertical_analyzer}, with the analyzer fixed in a vertical orientation relative to the sample surface, i.e., $Z(XY)\Bar{Z}$. This setup is the inverse of the configuration presented in the main text. While reversing the analyzer alters the pattern, the overall variation remains consistent.

\newpage

\textbf{Excitation wavelength-dependence.}

\begin{figure}[H]
	\centering
	\includegraphics[width=0.95\textwidth]{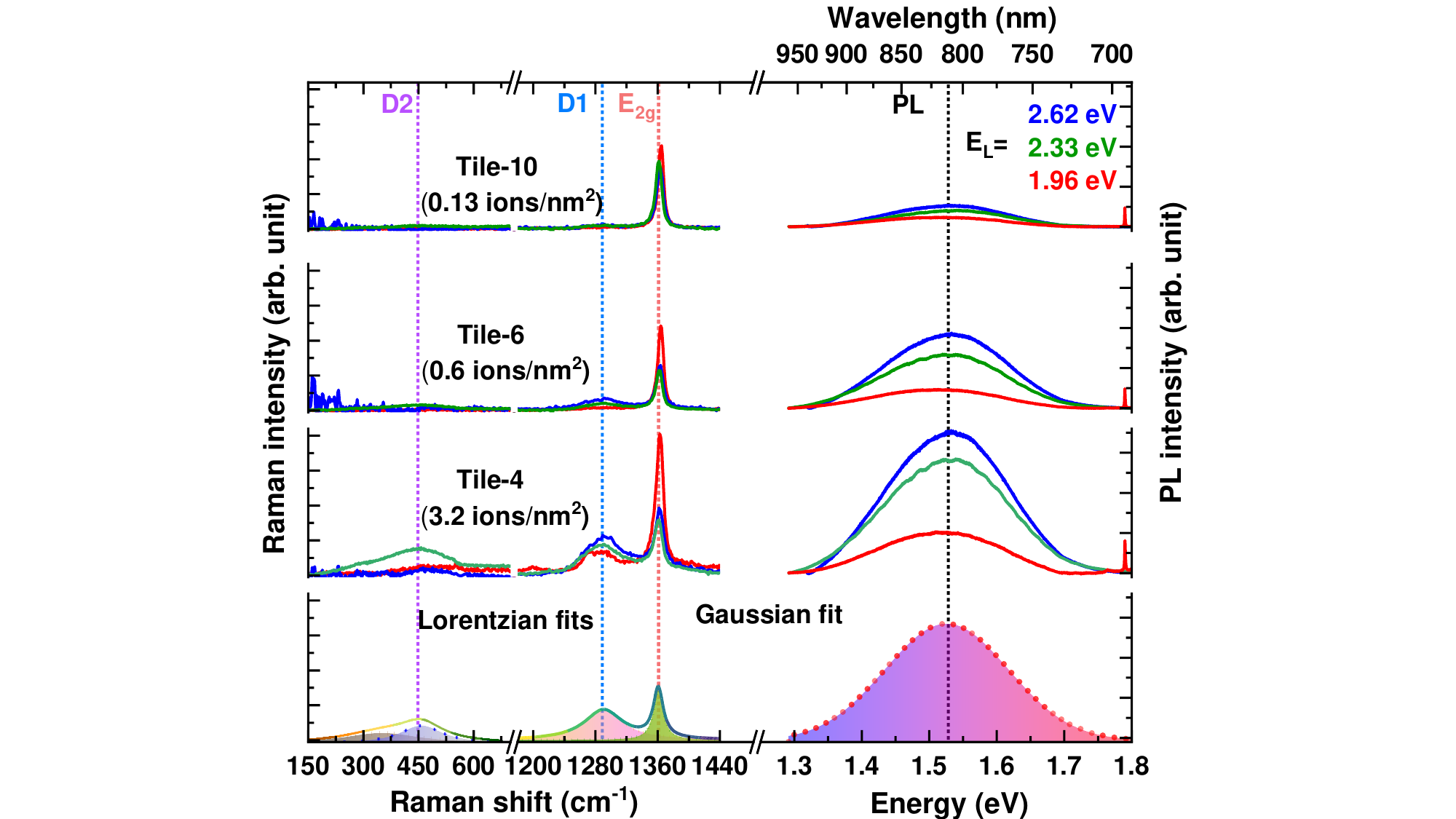}
	\caption{The dependence of background subtracted Raman and PL spectra on irradiation fluences of 3.2, 0.6, 0.13 ions/nm$^2$ along with different $E_\mathrm{L}$ values, 2.62 eV (blue), 2.33 eV (green) and 1.96 eV (red), is presented. The scale of all left-hand graphs is uniform, and similarly, all right-hand graphs share the same axis. The bottom panel illustrates the fitting of Raman and PL modes using Lorentzian and Gaussian functions, respectively. }
	\label{fig:Raman/PL_wavelength_wavelength}
\end{figure}

 \autoref{fig:Raman/PL_wavelength_wavelength} shows PL and Raman modes of Sample-2 for different excitation energies $E\mathrm{_L}$ of 2.62, 2.33, and 1.96 eV, shown in blue, green, and red, respectively. The measurements for three irradiation fluences (tiles) are presented with a vertical offset.
All left-hand graphs of \autoref{fig:Raman/PL_wavelength_wavelength} , representing background subtracted Raman intensity, share a common scale for easier comparison between irradiation fluences. Similarly, all right-hand graphs depict background subtracted PL intensity with a common scale.
The fittings of  Raman and PL spectra with Lorentzian and  Gaussian functions, respectively, are displayed in the bottom panel of \autoref{fig:Raman/PL_wavelength_wavelength}  for $E\mathrm{_L}= \SI{2.33}{\electronvolt}$.

The noticable dependence of PL on $E\mathrm{_L}$ can be explained by two key factors. First, as $E\mathrm{_L}$ diverges from the absorption energy to lower energies, a significant decrease in the light-matter interaction is observed. 
The absorption energy, identified as approximately 2.6 eV (476 nm) from PL excitation measurements,${}^2$ indicates a region of maximum optical response.  
This result underscores the significant role of this spectral region in the efficient absorption and subsequent PL processes within the material.
The second factor is that excitation with $E\mathrm{_L}= \SI{2.33}{\electronvolt}$ may also introduce photo-activated electrons from the hBN valance band, promoting the ionization process ${\mathrm{V_B}^0 + e^- \rightarrow \mathrm{V_B}^-}$, thereby enhancing PL.${}^3$ 
This might explain the strong signal at 2.33 eV even though it is away from its absorption energy. Thus, the combined effects of absorption energy and ionization processes govern the dependence of PL on $E_\mathrm{L}$.
The Raman modes also show dependence on $E_\mathrm{L}$.  For pure Raman scattering, the peak intensity of the modes is expected to increase with decreasing $E_\mathrm{L}$.
We find that only the \eg Raman mode behaves as anticipated. 
Remarkably, the D1 Raman mode displays a similar behavior as the PL peak pointing to a strong correlation to defect density. The contradiction to the expected behaviour could be resolved by the assumption of an increase in \vb defects caused by an ionization process. 
This explain the increased intensity of the defect-related signals PL and D1.

\newpage 
\textbf{Raman and PL linewidth dependence on irradiation fluence.} 
\begin{figure}[H]
	\centering
	\includegraphics[width=0.95\textwidth]{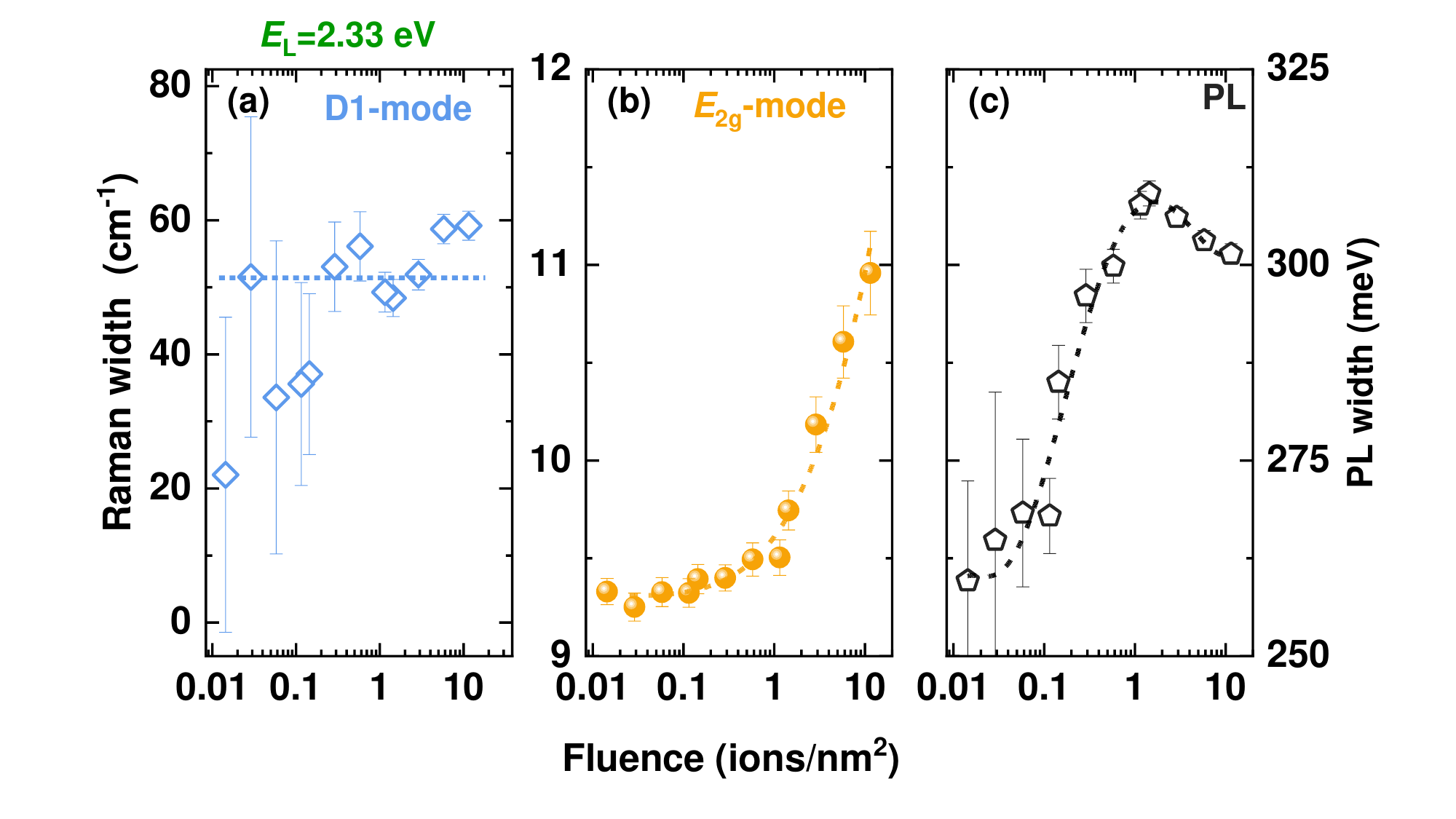}
	\caption{Variation in peak width: (a) D1 mode, (b) \eg  mode, and (c) PL. As the crystallinity of hBN is impacted by the irradiation fluence, the Raman width $\Gamma$ of the \eg  mode changes significantly. A lower $\Gamma$ at lower fluences indicates higher quality of the hBN sample. }
	\label{fig:Raman width}
\end{figure}

The main text explains how the integrated areas of the D1 and \eg Raman modes, and PL signal vary with different nitrogen fluences. Additionally, \autoref{fig:Raman width}  illustrates that the corresponding linewidths are also influenced by defects. As the irradiation fluence increases, the \eg-mode linewidth becomes wider, indicating a gradual increase in structural deformation. A narrower \eg -mode indicates fewer defects in the crystal. Unlike the D1 Raman mode, the PL linewidth broadens with increasing fluence, which can be explained by the higher number of defects contributing to radiative recombination, thereby increasing PL intensity. However, the variability in the electric dipole moments associated with individual defects results in a broader PL linewidth at higher fluences.

\newpage

\begin{figure}[H]
	\centering
	\includegraphics[width=0.99\textwidth]{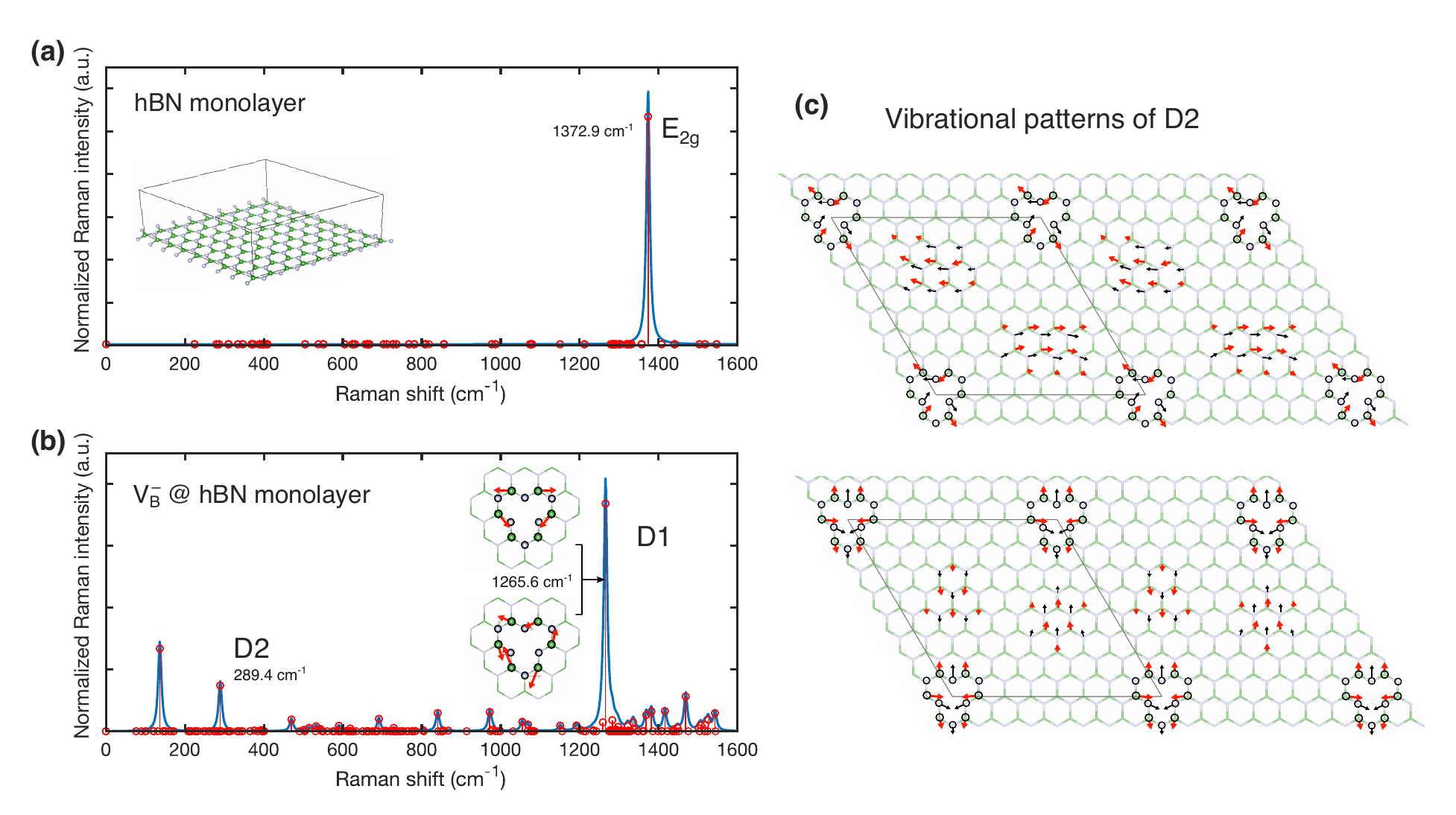}
	\caption{Raman activities for vibrational modes (red) and the corresponding Raman spectra calculated with DFT for a monolayer of pristine hBN containing 128 atoms (8×8 unit cells) (a) and the monolayer of the same size containing the \vb defect (b). Vibrational patterns of defect-induced modes D1 and D2 are shown in the inset of (b) and in (c). Red and black arrows indicate displacement amplitudes of boron and nitrogen atoms, respectively.}
	\label{fig:DFT_8x8layer_full}
\end{figure}

\begin{figure}[H]
	\centering
	\includegraphics[width=0.7\textwidth]{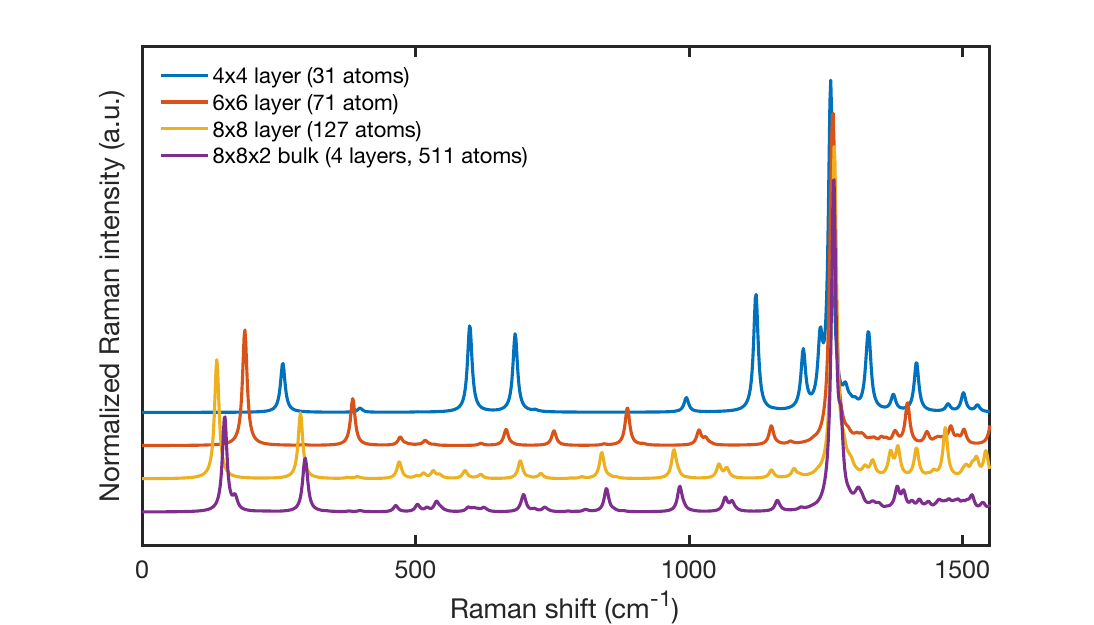}
	\caption{DFT-calculated Raman spectra obtained for different supercell sizes of defect-containing monolayers of 4×4 unit cells, 6×6 unit cells, and 8×8 unit cells, and bulk containing 8×8×2 unit cells / 4 monolayers.}
	\label{fig:DFT_size}
\end{figure}

\textbf{Calculated Raman modes.} 
In addition to the DFT results presented in the main text, \autoref{fig:DFT_8x8layer_full} provides the complete calculated spectra for an 8×8 pristine and \vb-containing monolayer. The vibrational patterns of the D2 mode, shown in \autoref{fig:DFT_8x8layer_full} (c), emphasize its quasi-local nature, with vibrations extending beyond the immediate defect region. \autoref{fig:DFT_size} demonstrates the influence of supercell size on the calculated Raman spectra. The D1 mode is unaffected by system size, confirming its localized character, while the D2 mode exhibits a strong dependence on supercell size, reflecting its sensitivity to defect concentration.

\begin{figure} 
	\centering
	\includegraphics[width=0.99\textwidth]{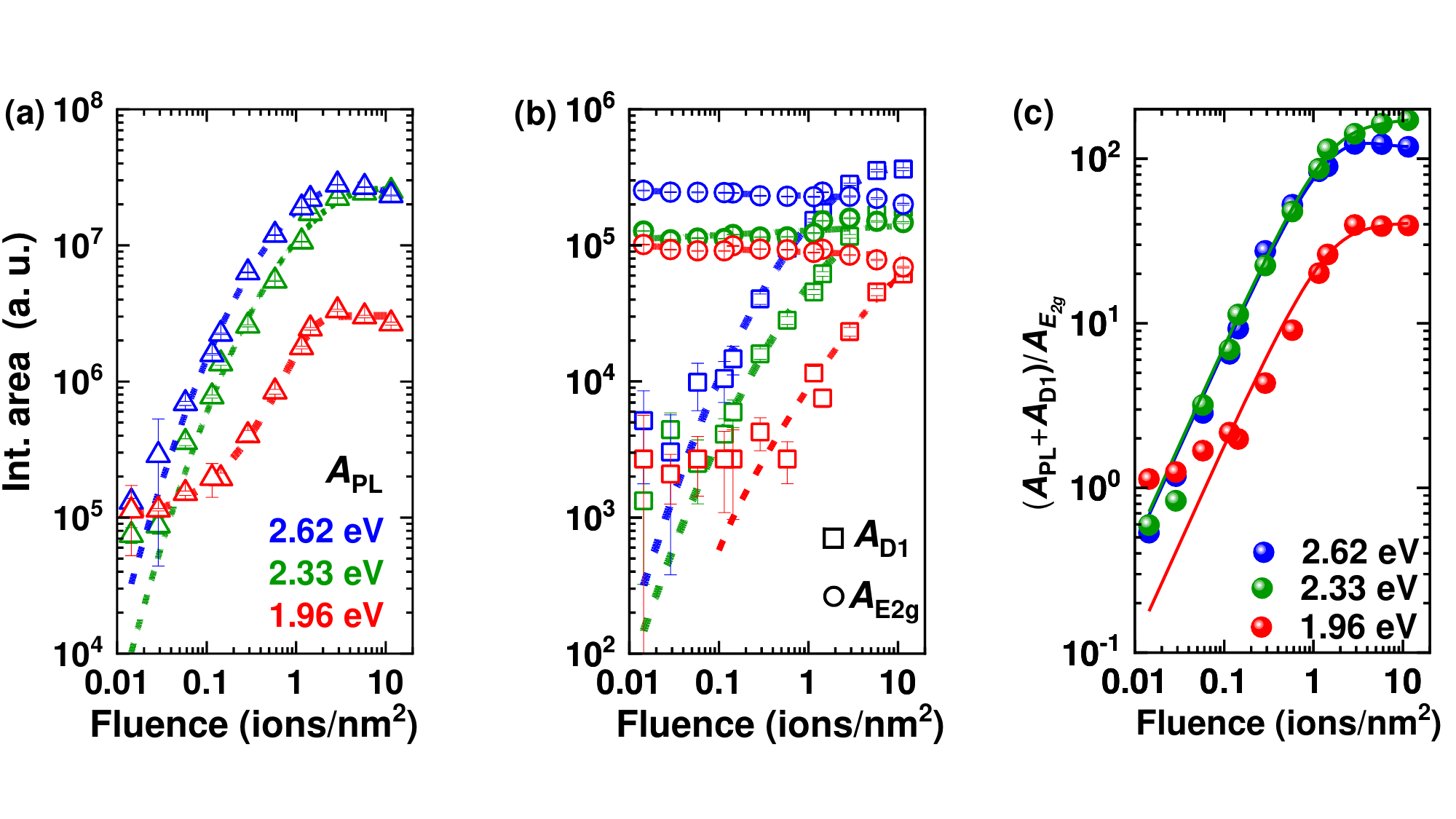}
	\caption{The variation of obtained  $A_\mathrm{PL}$ and $A_\mathrm{D1}$ and  $A_{E_\mathrm{2g}}$ for three  $E_\mathrm{L}$ are shown in (a) and (b), respectively. The ratio $(A_\mathrm{D1}+A_\mathrm{PL})/A_{E_\mathrm{2g}}$ is presented in (c).}
	\label{fig:intensity_EL}
\end{figure}

\textbf{Correlation of Raman mode and PL signals.}
\autoref{fig:intensity_EL} (a) and (b) presents the variation of $A_\mathrm{PL}$ and $A_\mathrm{D1}$ and  $A_{E_\mathrm{2g}}$ respectively for different $E_\mathrm{L}$, while the ratio $(A_\mathrm{D1}+A_\mathrm{PL})/A_{E_\mathrm{2g}}$ is shown in (c).

\begin{figure} 
	\centering
	\includegraphics[width=0.8\textwidth]{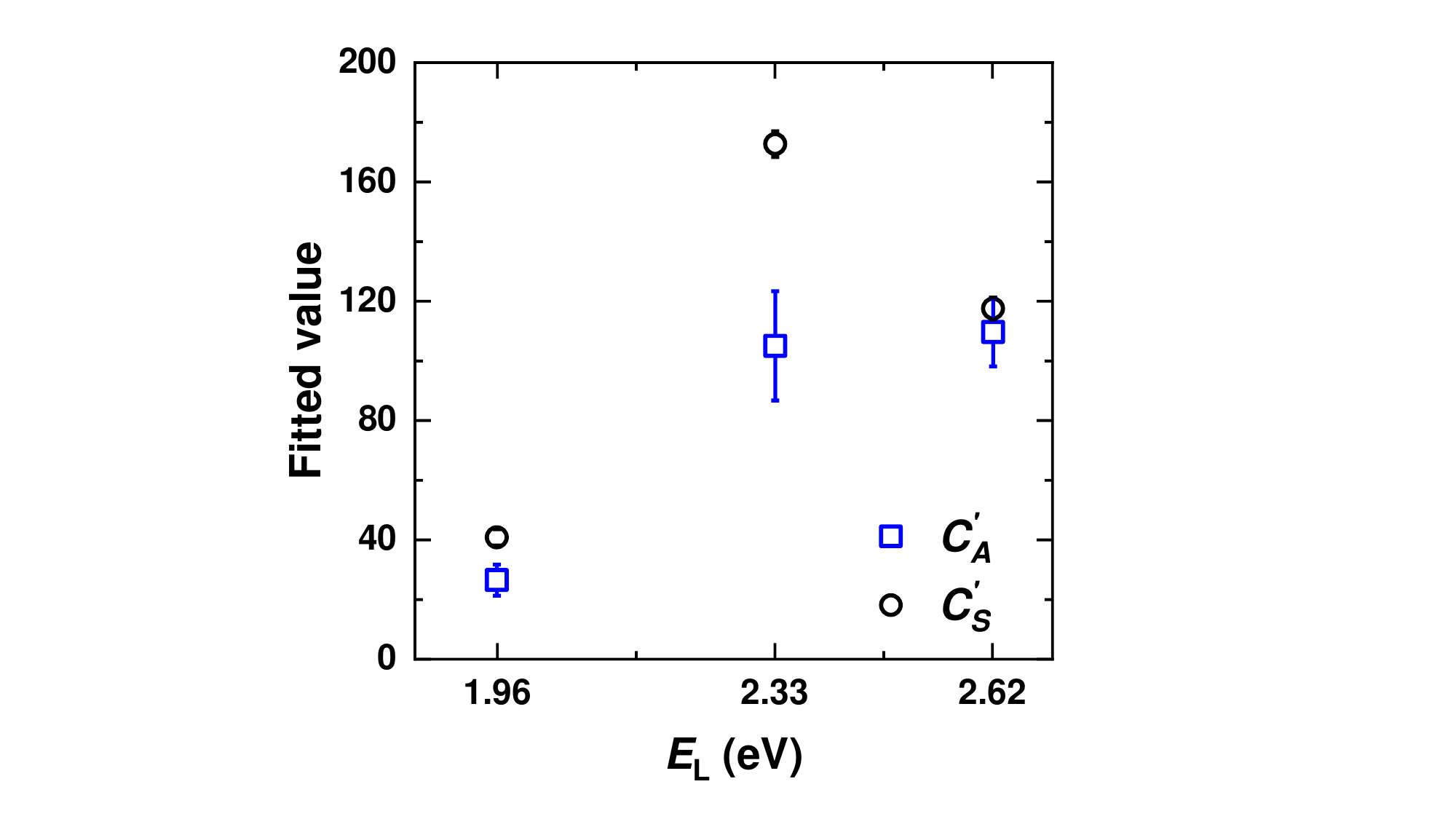}
	\caption{The variation of obtained $C_A^{\prime}$ and $C_S^{\prime}$ values by fitting of Fig. 4 (b) with \autoref{eq:eqn-2} is shown against $E_L$. 
    }
	\label{fig:variation with EL}
\end{figure}

\newpage 
\textbf{Wavelength dependence of fit parameters.}
The dependencies of the two fitting parameters, $C_A^{\prime}$ and  $C_S^{\prime}$, extracted by fitting \autoref{eq:eqn-2}, are shown in  \autoref{fig:variation with EL}, as functions of excitation energy. Here, $C_A^{\prime}$ defines the maximum possible value of $(A_\mathrm{D1}+A_\mathrm{PL})/A_{E_\mathrm{2g}}$. Both parameters vary with E$_L$. 

\newpage

\begin{figure}
	\centering
	\includegraphics[width=0.99\textwidth]{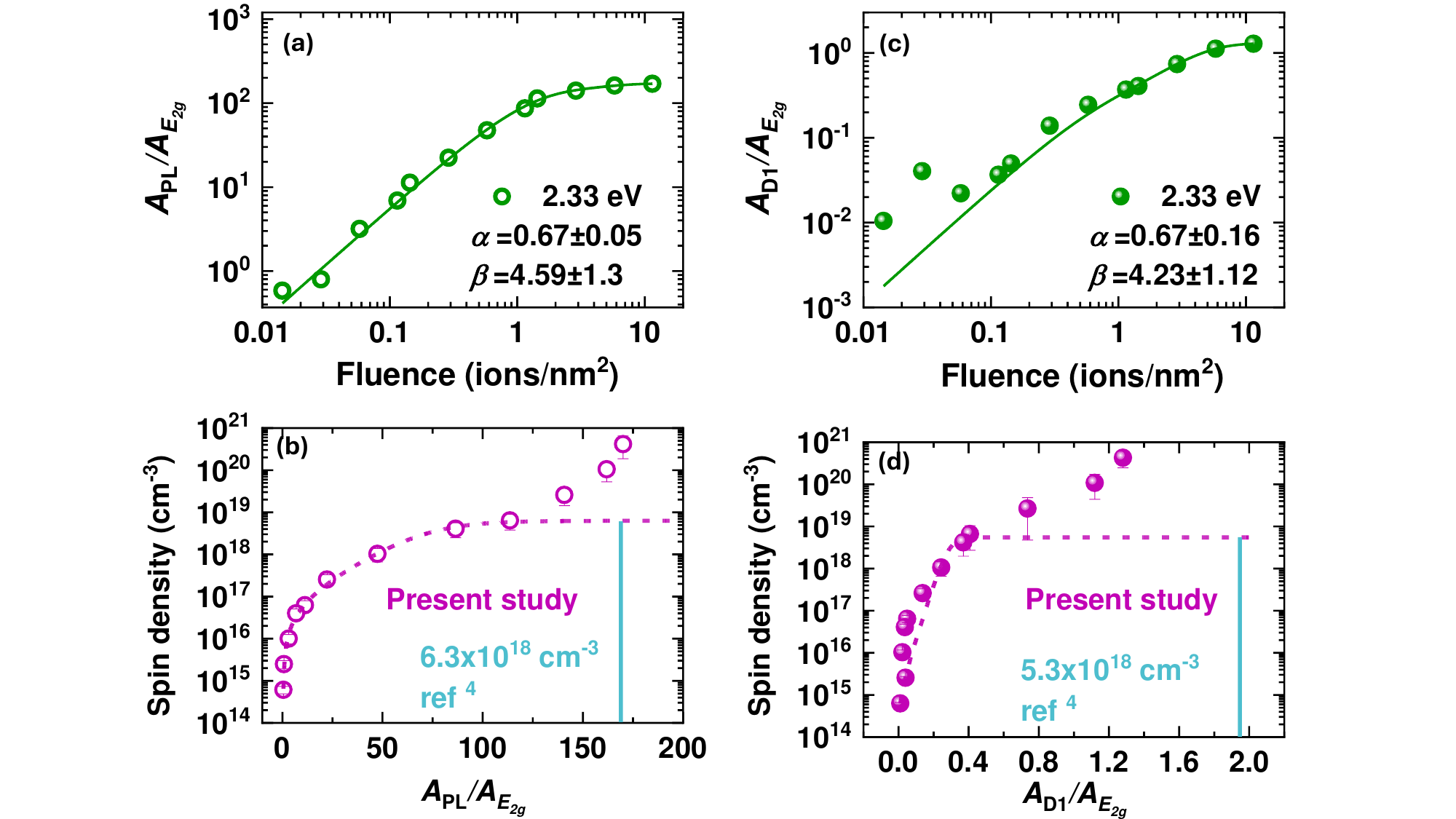}
	\caption{The variation of A$_\mathrm{PL}$/$A_{E_\mathrm{2g}}$ and A$_\mathrm{D1}$/$A_{E_\mathrm{2g}}$ with irradiation fluence is shown in (a) and (c), respectively. 
    The solid line represents the fit using \autoref{eq:eqn-2}  from the main text. Corresponding spin density calculations are presented in (b) and (d), respectively.}
	\label{fig:calculation_Raman_PL}
\end{figure}

\textbf{Spin defect density determination using $A_\mathrm{PL}/A_\mathrm{E_{2g}}$ and $A_\mathrm{D1}/A_\mathrm{E_{2g}}$.} 
In contrast to the main text, where the spin density is calculated considering three parameters -- PL and D1 with respect to \eg, here we present the same calculation using only two parameters. The ratios $A_\mathrm{PL}/A_\mathrm{E_{2g}}$ and $A_\mathrm{D1}/A_\mathrm{E_{2g}}$ are shown in \autoref{fig:calculation_Raman_PL} (a) and \autoref{fig:calculation_Raman_PL} (c), respectively for Sample-1. The corresponding calculated densities are shown in \autoref{fig:calculation_Raman_PL} (b) and \autoref{fig:calculation_Raman_PL}(d). 

\newpage

\begin{figure}[H]
	\centering
	\includegraphics[width=0.95\textwidth]{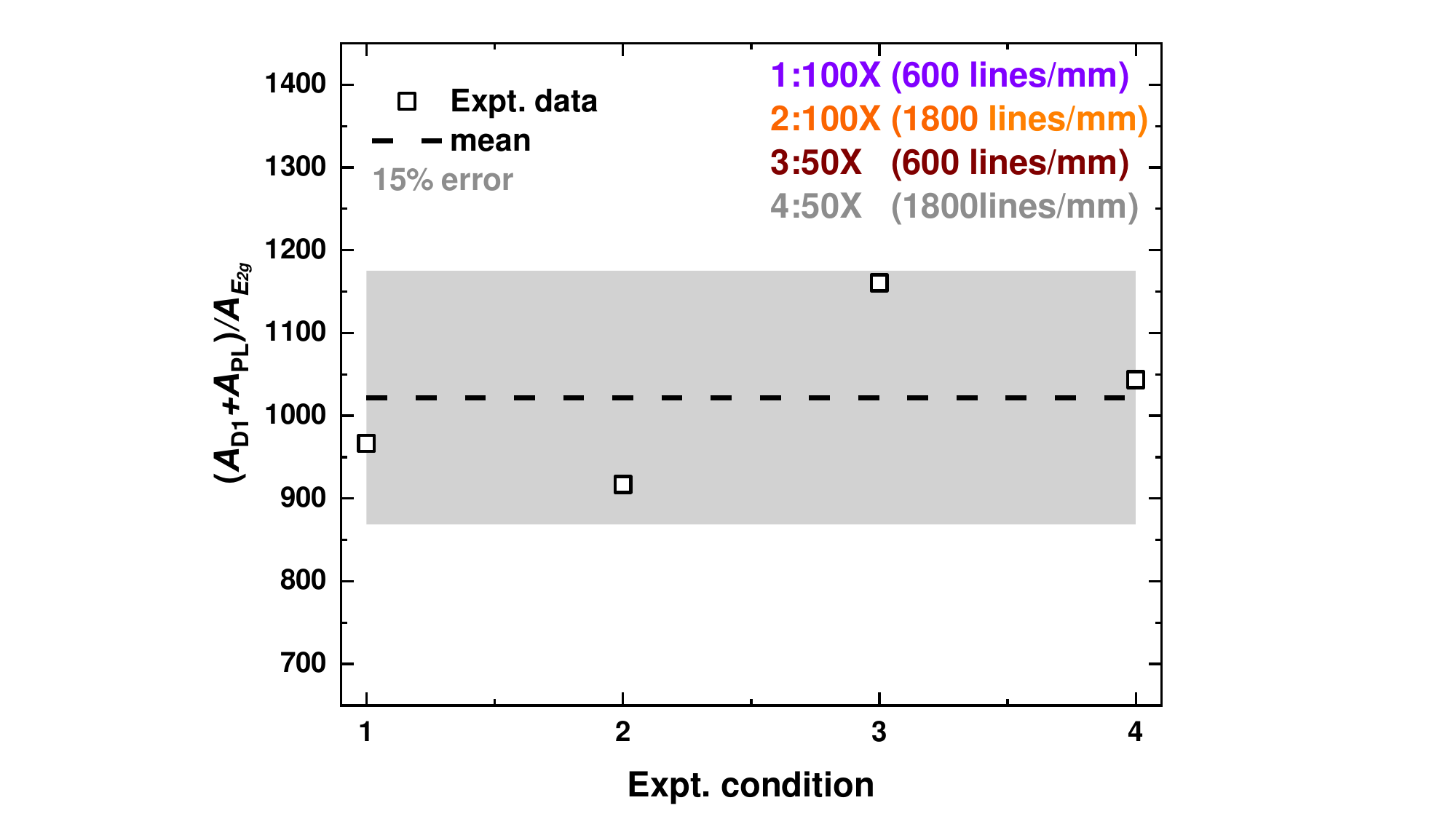}
	\caption{Variation of $(A_\mathrm{D1}+A_\mathrm{PL})/A_\mathrm{E_{2g}}$ measured with different combination of objectives and grating.}
	\label{fig:expt condition}
\end{figure}

\textbf{Effect of instrument parameters.}
The variation of $(A_\mathrm{D1}+A_\mathrm{PL})/A_{E_\mathrm{2g}}$ of Sample-2 is shown in \autoref{fig:expt condition}. This variation reflects how the choice of optical components and experimental setups can influence spectral measurements. Despite these variations, the $(A_\mathrm{D1}+A_\mathrm{PL})/A_\mathrm{E_{2g}}$ values changed by less than 15$\%$, demonstrating the robustness of this parameter for calculating spin density and validating its reliability across different experimental conditions.

\newpage

\begin{figure}[H]
	\centering
	\includegraphics[width=0.99\textwidth]{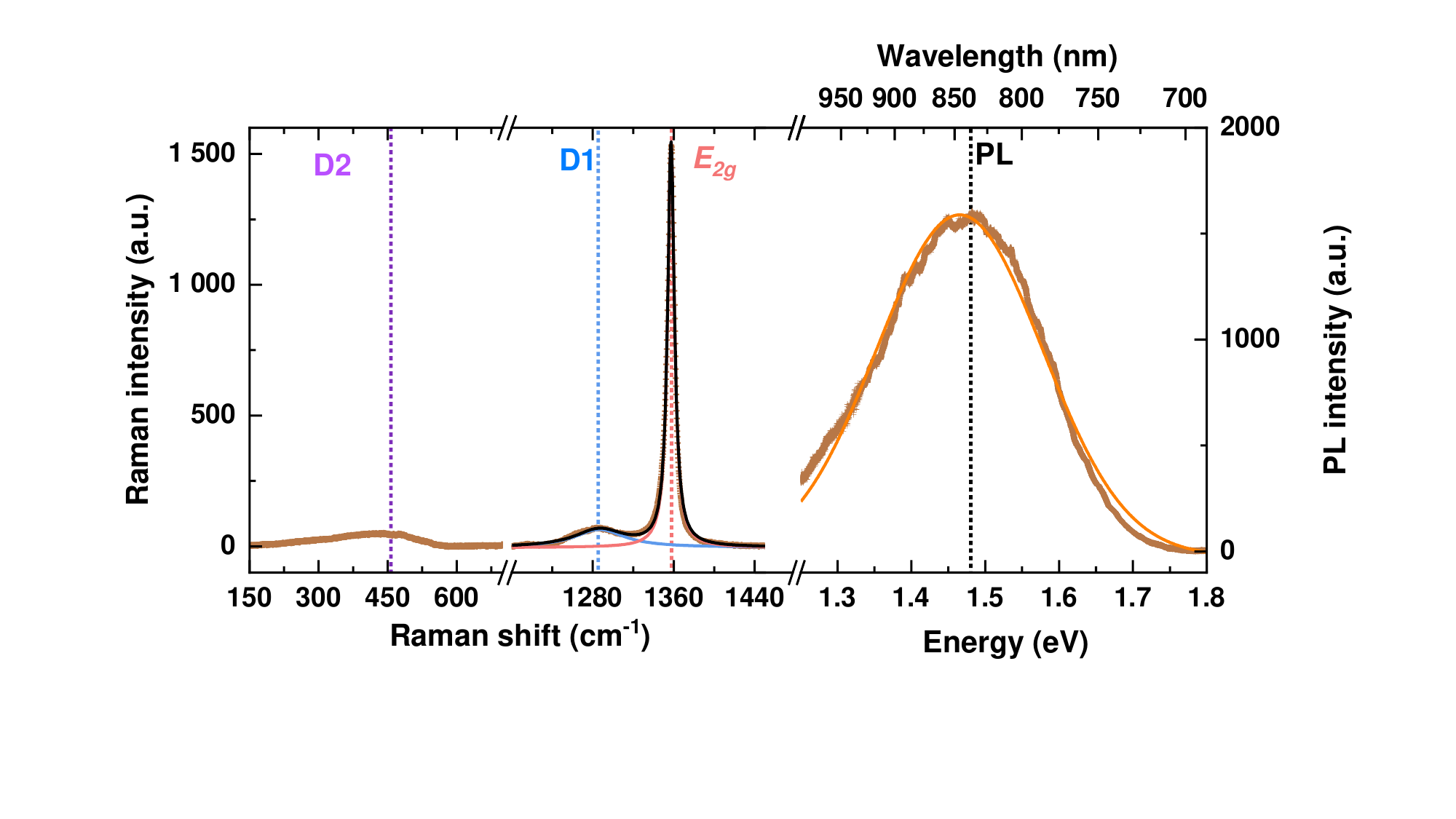}
	\caption{The value of $(A_\mathrm{D1}+A_\mathrm{PL})/A_\mathrm{E_{2g}}$ is calculated as 191 from the integrated area of fitted data. Spectral information adopted from reference${}^4$ with permission.} 
	\label{fig:Raman_PL_ref}
\end{figure}

\textbf{Spectral analysis of previously reported work.} 
For comparison, a spectrum from ref $^4$ for a neutron irradiated crystal is presented in \autoref{fig:Raman_PL_ref} together with spectral fits. The parameter $(A_\mathrm{D1}+A_\mathrm{PL})/A_{E_\mathrm{2g}}$ yields a value of 191. With this value we estimate the spin density, as shown in \autoref{fig:Fitting} (d) of the main text with an asterisk.

\newpage

\newpage

\noindent\textbf{References}\\
\indent(1) Kotakoski, J.; Jin, C. H.; Lehtinen, O.; Suenaga, K.; Krasheninnikov, A. V. Electron knock-on damage in hexagonal boron nitride monolayers. \textit{Phys. Rev. B} \textbf{2010}, \textit{82}, 11.

(2) Baber, S.; Malein, R. N. E.; Khatri, P.; Keatley, P. S.; Guo, S.; Withers, F.; Ramsay, A. J.; Luxmoore, I. J. Excited state spectroscopy of boron vacancy defects in hexagonal boron nitride using time-resolved optically detected magnetic resonance. \textit{Nano Lett.} \textbf{2021}, \textit{22}, 461–467.

(3) Gale, A.; Scognamiglio, D.; Zhigulin, I.; Whitefield, B.; Kianinia, M.; Aharonovich, I.; Toth, M. Manipulating the charge state of spin defects in hexagonal boron nitride. \textit{Nano Lett.} \textbf{2023}, \textit{23}, 6141–6147.

(4) Gottscholl, A.; Diez, M.; Soltamov, V.; Kasper, C.; Krauße, D.; Sperlich, A.; Kianinia, M.; Bradac, C.; Aharonovich, I.; Dyakonov, V. Spin defects in hBN as promising temperature, pressure and magnetic field quantum sensors. \textit{Nat. Commun.} \textbf{2021}, \textit{12}, 4480.


\end{document}